\documentclass[10pt,journal,compsoc,authordraft]{IEEEtran}

\usepackage{verbatim}
\usepackage{multirow}
\usepackage{booktabs}
\usepackage{bigstrut}
\usepackage{hyperref}
\usepackage{xcolor}
\usepackage{graphicx}
\usepackage{amsmath}
\usepackage[ruled]{algorithm2e}
\usepackage{array}
\usepackage{subcaption}
\usepackage{sidecap}
\usepackage{wrapfig}
\usepackage{etoolbox}
\usepackage{adjustbox}
\usepackage{diagbox}
\usepackage[final]{pdfpages}
\usepackage{booktabs}

\newcommand{\PreserveBackslash}[1]{\let\temp=\\#1\let\\=\temp}
\newcolumntype{C}[1]{>{\PreserveBackslash\centering}p{#1}}
\newcolumntype{R}[1]{>{\PreserveBackslash\raggedleft}p{#1}}
\newcolumntype{L}[1]{>{\PreserveBackslash\raggedright}p{#1}}

\newcommand{\design}{\texttt{CStream}\xspace}

\newcommand{\compress}{compression\xspace}

\usepackage{hyphenat}
\usepackage[skip=7pt]{caption}

\usepackage{enumitem}

\newcommand{\xianzhi}[1]{{\textcolor{purple}{#1}}}	

\begin{document}

\IEEEtitleabstractindextext{%
\begin{abstract}
In the burgeoning realm of Internet of Things (IoT) applications on edge devices, data stream compression has become increasingly pertinent. The integration of added compression overhead and limited hardware resources on these devices calls for a nuanced software-hardware co-design. This paper introduces \design, a pioneering framework crafted for parallelizing stream compression on multicore edge devices. \design grapples with the distinct challenges of delivering a high compression ratio, high throughput, low latency, and low energy consumption.
Notably, \design distinguishes itself by accommodating an array of stream compression algorithms, a variety of hardware architectures and configurations, and an innovative set of parallelization strategies, some of which are proposed herein for the first time. Our evaluation showcases the efficacy of a thoughtful co-design involving a lossy compression algorithm, asymmetric multicore processors, and our novel, hardware-conscious parallelization strategies. This approach achieves a $2.8x$  compression ratio with only marginal information loss, $4.3x$ throughput, $65\%$ latency reduction and $89\%$ energy consumption reduction, compared to designs lacking such strategic integration.
\end{abstract}

}
\title{CStream: Parallel Data Stream Compression on Multicore Edge Devices}
\author{
Xianzhi Zeng, Shuhao Zhang
\IEEEcompsocitemizethanks{
    \IEEEcompsocthanksitem Xianzhi Zeng and Shuhao Zhang are with the Singapore University of Technology and Design.
}\thanks{}}
\maketitle

\section{Introduction}
\label{sec::introduction}
\begin{figure}[t]
    \centering
    \includegraphics*[width=0.5\textwidth]{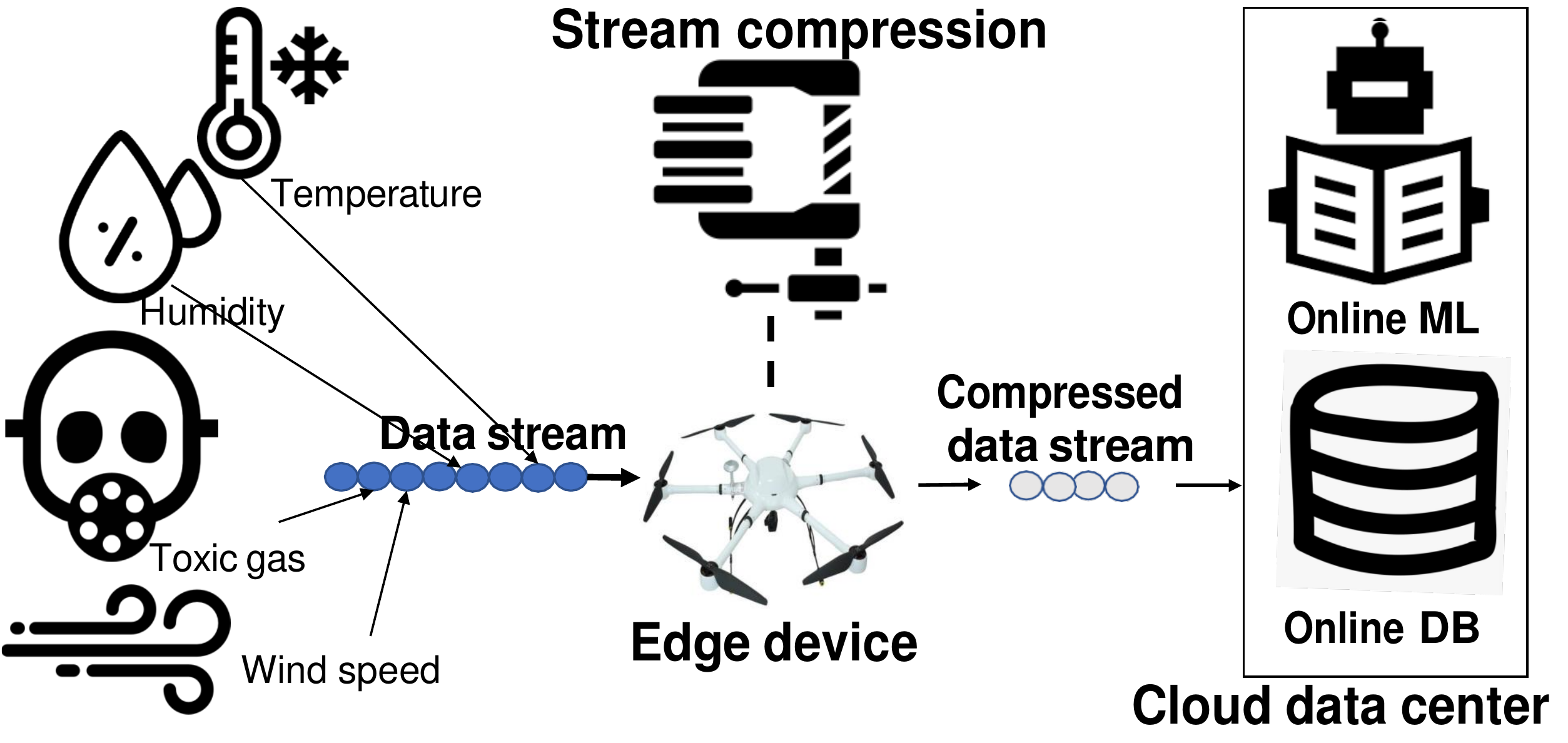}
    \caption{Application of stream compression in real-time data gathering conducted by a patrol drone in environments inaccessible to humans.}
    \label{fig:sc_comp_exp}   
\end{figure}

With the rise of Internet of Things (IoT) applications, the need for efficient data processing in edge devices, especially data stream compression, has become a pivotal research problem~\cite{comp_tersecades,streamZip}. Figure~\ref{fig:sc_comp_exp} illustrates an IoT use case~\cite{zeng2023hardware} wherein stream compression in multicore edge devices is highly desirable. Real-time data streams (e.g., toxic gas, temperature) from a multitude of IoT sensors in hazardous areas are incessantly gathered by patrol drones, functioning as edge devices, with limited memory and battery power. 
To reduce transmission overhead, these patrol drones, equipped with multicore processors, act as multicore edge devices that compress the input streams~\cite{CStreamICDE} before passing them to downstream online IoT analytic tasks, such as online aggregation~\cite{streamZip}, and online machine learning~\cite{ml_li2022camel} in the cloud.

Parallelizing stream compression on multicore edge devices, such as the wireless patrol drones in Figure~\ref{fig:sc_comp_exp}, is mandatory to meet the strict high-throughput processing requirements. However, achieving this in the resource-constrained environment of multicore edge devices is a non-trivial task. It involves striking a delicate balance between often conflicting requirements such as low energy consumption~\cite{nebulastream}, high compression ratio~\cite{ml_li2022camel}, and tolerable information loss~\cite{comp_tersecades}. While data stream compression is a well-studied problem, the specific context of multicore edge devices adds a new dimension to it. In particular, none provide a comprehensive answer to our central question: \begin{quote}
\textit{How can stream compression be optimally implemented on multicore edge devices for IoT applications?}
\end{quote}

This paper introduces \design, an innovative framework specifically engineered for parallelizing stream compression on multicore edge devices. \design systematically navigates the expansive design space, meticulously balancing various factors such as \emph{compression ratios}, \emph{compression speeds}, and \emph{energy consumption}. It offers a flexible, adaptive, and robust solution that pushes the boundaries of stream compression on multicore edge devices, exploiting a versatile software-hardware co-design approach~\cite{IoT_survey2019,amp_iot}.

Firstly, \design supports a wide array of stream compression algorithms, each possessing unique strengths and trade-offs. These algorithms span from conventional lossless compression to cutting-edge lossy algorithms, encapsulating both stateful and stateless variations, and those with or without byte alignment.

Secondly, \design is conscientious of hardware differences, demonstrating adaptability across diverse hardware architectures and configurations. It is capable of optimizing its functionality according to the specificities of various multicore processors, accommodating RISC or CISC architectures, and adjusting to varying word lengths and core numbers. This adaptability allows \design to optimally harness the hardware resources available on different edge devices, thereby maximizing compression speed while minimizing energy consumption.

Thirdly, \design introduces a novel set of parallelization strategies, some of which are proposed for the first time. These strategies account for various factors such as execution strategies, state-sharing implementation, and scheduling strategies, offering fine-grained control over the parallelization of stream compression. With these innovative parallelization strategies, \design is capable of better distributing the compression workload across multiple cores, thereby enhancing throughput and reducing latency.

Our primary contributions are summarized as follows:

\begin{itemize}
\item \emph{First}, \design offers an extensive set of stream compression algorithms, with a particular focus on \emph{lossy stream compression} algorithms. These algorithms strike a balance between high compression ratios (ranging from 2.0 to 8.5) and minimal information loss (less than 5\%), enabling \design to cater to a wide spectrum of application requirements.
\item \emph{Second}, \design is engineered to operate efficiently on asymmetric multicore processors with RISC architecture and 64-bit word length. This results in impressive performance improvements, specifically, up to 59\% reduction in processing time and up to 69\% reduction in energy consumption when compared to traditional multicore processors.
\item \emph{Third},  \design implements a range of hardware-conscious parallelization strategies. To begin with, it incorporates cache-aware micro-batching of tuples, which significantly improves throughput by up to 11 times. Next, in terms of state management, it opts for private dictionaries for each thread instead of a shared state. This optimization significantly reduces energy consumption and enhances throughput without adversely affecting the compression ratio. 
\item \emph{Lastly}, \design adopts asymmetry-aware workload scheduling~\cite{CStreamICDE}. This not only reduces energy consumption by about 50\% but also optimizes resource utilization by leveraging the unique capabilities of different cores. These strategies, collectively, make \design an efficient and effective tool for stream compression on heterogeneous multi-core systems.
\end{itemize}

In demonstrating the efficacy of \design, we conducted a comprehensive evaluation with five real-world and one synthetic datasets, featuring diverse workload characteristics. Our results confirm \design's superior performance over traditional approaches, demonstrating its effectiveness in the challenging environment of multicore edge devices (e.g., RK3399~\cite{amp_rk3399}, H2+\cite{amp_h2+}, and Z8350\cite{amp_z8350}). Our observations underscore the value of thoughtful co-design in achieving optimal stream compression on edge devices. We highlight the potential of \emph{lossy stream compression} algorithms, asymmetric multicore processors, and hardware-conscious parallelization strategies. These strategic integrations lead to notable improvements in the compression ratio, throughput, and energy efficiency. With these contributions, we envision \design to be an indispensable tool for researchers and practitioners aiming to achieve efficient data stream compression for IoT applications.
 
\textbf{Organization.}
The rest of this paper is laid out as follows:
Section~\ref{sec:preliminaries} offers an overview of stream compression and underscores the unique challenges associated with its application at the edge.
Section~\ref{sec:co-design} delves into the co-design spaces of stream compression at the edge, underlining the architecture and implementation of \design.
In Section~\ref{sec:methodology}, we present the methodology used to evaluate \design.
Section~\ref{sec:evaluation} reports on our experimental evaluation of \design, highlighting its performance across various co-design spaces.
Section~\ref{sec:related} situates our work within the existing body of literature, emphasizing how \design advances the current state of the art in stream compression for multicore edge devices.
Section~\ref{sec:conclusion} concludes the paper, reflecting on the contributions of \design and offering directions for future research in this domain.
\section{Background}
\label{sec:preliminaries}
In this section, we introduce the basic concepts of stream compression at the edge.

\subsection{Overview of Stream Compression}
\label{subsec:overview_sc}
In our definition, a stream tuple, denoted as $x_{t}=v$, consists of an arrival timestamp $t$ to the processing system (e.g., a compressor), and $v$, the content to be compressed. We define a list of tuples chronologically arriving at the system as an ~\emph{input stream}. Stream compression essentially performs the task of \emph{continuously compressing the input stream into an output stream with fewer data footprints}. To describe the relative size between the loaded (i.e., input) and compressed (i.e., output) streaming data, we define \emph{compression ratio} as $compression \ ratio=\frac{loaded \ data \ size}{compressed \ data \ size}$.

Stream compression possesses three major properties stemming from the nature of the continuously arriving data stream. First, the stream is incremental, meaning that compression on the \emph{current tuple $x_\tau$} (i.e., the tuple arriving most recently at time $\tau$) can rely on either 1) itself or 2) the \emph{past tuples} (i.e., those arriving earlier than the current tuple with timestamp $t<\tau$). However, it has no reference to the \emph{future tuples} which haven't arrived yet with timestamp $t>\tau$.

Second, the stream is infinite. Therefore, the proper utilization of cache and memory becomes crucial. Third, the stream is characterized by a large volume and high rate, necessitating high-throughput compression. These distinct properties set stream compression apart from database compression~\cite{db_comp0}, time series compression~\cite{blalock2018sprintz}, and file compression~\cite{comp_zlib}, where all data is ready before conducting the compression.

\subsection{Stream Compression on Edges}
\label{subsec:sc_at_edge}
The emergence of IoT has led to an explosion in data collection, storage, and processing demand at the edge. Here, we briefly introduce the data properties and performance demand of stream compression on edges.

\textbf{IoT Data.}
IoT, being a common application scenario for edge computing, produces diverse data streams at the edge~\cite{lu2020adaptively,nebulastream}. The source(s) of these data streams could be singular (e.g., an ECG sensor~\cite{comp_mit_ecg}) or multiple (e.g., distributed game servers~\cite{sp_rovio}). Additionally, the tuple content may be a plain value (e.g., unsigned integer~\cite{comp_mit_ecg}), binary structured (e.g., the $<key,value>$ pair~\cite{ds_stock}), or textual structured (e.g., in XML format~\cite{ds_sensor}). The arrival pattern and compressibility of the data stream can also greatly vary. For instance, an anti-Covid19 tracking system~\cite{iot_traceTogether} at a mall may generate data streams more densely during peak hours. Due to this versatility, efficient stream compression is highly context-dependent.

\textbf{Performance Demand.}
Given the \emph{13 V's challenges~\cite{IoT_survey2019}} of IoT, stream compression at the edge needs to meet several critical demands:

\begin{itemize}
\item \emph{High compression ratio:} Reducing the data footprint in the output stream is a key reason for using stream compression at the edge. With the data generated at the edge being nearly infinite and growing to ZB level per year recently~\cite{iot_data_scale}, and considering the limited memory capacity and communication bandwidth~\cite{nebulastream} on edge devices, a high compression ratio becomes necessary.

\item \emph{High throughput:} High throughput is desirable in stream analytics, not only at the data center~\cite{sp_briskstream,sp_coprocessing,comp_tersecades}, but also at the edge. The high-rate data streams at the edge, collected from massive device links~\cite{ukil2015iot,nebulastream}, necessitate the capability to compress as much data as possible within a given unit of time.

\item \emph{Low energy consumption:} Energy budget is particularly constrained at the edge, and the devices, often far from a stable and constant power supply, are expected to function as long as possible~\cite{edge_survey}. Thus, energy consumption should be minimized.

\item \emph{Low latency:} In many IoT applications, it's crucial to process data and make decisions in real time. Hence, low-latency stream compression is important to ensure timely response and decision-making at the edge.

\end{itemize}

Meeting these demands simultaneously is a challenging task. For instance, increasing parallelism and allocating more processor cores can achieve higher throughput, but this approach will also increase energy consumption. Therefore, this study aims to reveal the complex relationships between compression ratio, throughput, energy consumption, and latency of stream compression at the edge. We strive to offer comprehensive guidance for achieving satisfying trade-offs among these factors and also explore potential optimal approaches.
\section{CStream: Software-Hardware Co-Design of Stream Compression}
\label{sec:co-design}
Recognizing the unique challenges in edge computing environments, we introduce \design, a comprehensive software-hardware co-designed stream compression framework. \design stands out by accommodating an array of stream compression algorithms, a variety of hardware architectures and configurations, and an innovative set of parallelization strategies. Furthermore, it presents novel concepts, such as asymmetric-aware scheduling, for the first time. By leveraging the features of an advanced stateful compression algorithm and the inherent characteristics of asymmetric multicore platforms, \design efficiently handles diverse data types while meeting hardware resource constraints and maintaining energy efficiency. This section elaborates on the design and components of \design.

\begin{table}[]
\centering
\caption{Summary of studied compression algorithms}
\label{tab:algo_list}
\resizebox{0.89\columnwidth}{!}{%
\begin{tabular}{|p{2.8cm}|p{1.3cm}|p{1.9cm}|p{1.3cm}|}
\hline
\textbf{Algorithm Name}                                                      & \textbf{Fidelity}              & \textbf{State utilization}     & \textbf{Byte alignment} \\ \hline
LEB128-NUQ~\cite{comp_leb128,comp_nonuniform}       & lossy                 & stateless              & aligned        \\ \hline
ADPCM~\cite{comp_leb128,comp_nonuniform,comp_tersecades}       & lossy                 & stateful, value              & aligned        \\ \hline
UANUQ~\cite{comp_nonuniform,comp_elias}                              & lossy                 & stateless              & unaligned        \\ \hline
UAADPCM~\cite{comp_elias,comp_nonuniform,comp_tersecades}   & lossy                 & stateful, value              & unaligned        \\ \hline
LEB128~\cite{comp_leb128}                                           & lossless              & stateless             & aligned        \\ \hline
Delta-LEB128~\cite{comp_leb128,comp_tersecades}                      & lossless              & stateful, value              & aligned        \\ \hline
Tcomp32~\cite{comp_elias}                                           & lossless              & stateless             & unaligned      \\ \hline
Tdic32~\cite{comp_lz4,comp_elias}                                   & lossless              & stateful, dictionary              & unaligned      \\ \hline
RLE~\cite{wang2017experimental}                                   & lossless              & stateful, value             & aligned      \\ \hline
PLA~\cite{zhou2011dissemination,blalock2018sprintz}   & lossy              & stateful, model              & aligned      \\ \hline
\end{tabular}%
}
\end{table}
\subsection{Versatility of Compression Algorithms}
\label{subsec:compression_algorithm}
The heart of \design lies in its support for a wide variety of compression algorithms. It provides versatility to diverse IoT data types and optimizes compressibility while minimizing information loss. 
Compression algorithms play a significant role in the historical landscape of data processing, with numerous variations proposed since the 1950s~\cite{comp_huffman}. These algorithms typically target improvements in theoretical compressibility or reductions in compression overhead~\cite{comp_survey_2017,comp_survey_huffman}. In constructing \design, we encompassed ten representative compression algorithms that embody key features such as \emph{fidelity}, \emph{state utilization}, and \emph{alignment} (summarized in Table~\ref{tab:algo_list}).

\subsubsection{Fidelity}
\label{subsub:lossless_lossy}
One of the fundamental distinctions among compression algorithms is their fidelity: the degree to which the original data can be reconstructed from compressed data.

\textbf{Lossless Compression.}
Lossless compression is the most stringent form of fidelity, ensuring that the original data can be perfectly reconstructed from its compressed form. Under lossless compression, \design rearranges data into a more compact representation without altering its underlying meaning or losing any information. This principle is strictly bounded by theoretical limits, such as Shannon's entropy~\cite{comp_survey_huffman}, which serves as an upper bound on the effectiveness of lossless compression. One exemplar is the $Tcomp32$ algorithm~\cite{comp_elias} used in \design's lossless mode. It employs a lossless stream compression technique that suppresses leading zeros~\cite{comp_elias}, a form of bit-level null suppression~\cite{db_comp2}, and its compressibility limit aligns with Shannon's entropy. In this way, \design preserves the accuracy and fidelity of the original data even under compression.

\textbf{Lossy Compression.} 
Lossy compression, on the other hand, relaxes the fidelity requirement, accepting some loss of information to achieve higher compression ratios. Under lossy compression, \design selectively discards less significant information from the original data. The upper bound on compression effectiveness is then determined by the level of fidelity the system chooses to maintain. For instance, an unaligned non-uniform quantization (UANUQ~\cite{comp_nonuniform,comp_elias}) with 8 quantization bits would result in a higher compression ratio, and thus more information loss, than one using 12 quantization bits. \design's lossy mode is designed to strike a balance between high compression ratios and acceptable information loss, accommodating the wide range of fidelity requirements across different application scenarios at the edge.

\subsubsection{State Utilization}
\label{subsec:intro_state}
State utilization, which determines how a compression algorithm uses historical information, is another critical dimension in the design of \design. We divide it into stateless and stateful compression modes, each with unique advantages and best-use scenarios.

\textbf{Stateless Compression.}
In stateless compression, \design operates on each data item or tuple independently without reference to any past tuples. This approach essentially replicates a set of operations for each tuple, requiring no state management. This is particularly beneficial in scenarios where the relationships between consecutive tuples are insignificant or when reducing processing overhead is essential. Algorithm~\ref{alg:stateless} provides an overview of \design's stateless compression mode.

\begin{algorithm}
\caption{\design's Stateless Compression Mode}
\label{alg:stateless}
\scriptsize
\LinesNumbered 
\KwIn{input stream $inData$}
\KwOut{output stream $outData$}
\While{$inData$ is not stopped}
{
    (s0) load the tuple(s) from $inData$ \label{algo:stateless_S0} \;
    (s1) transform and find the \emph{compressible} parts\label{algo:stateless_S0}\;
    (s2) output compressed data to $outData$ \label{algo:stateless_S0}\;
}
\end{algorithm}

As a concrete example of stateless compression algorithm, we show the detailed implementation of Android-Dex's $LEB128$~\cite{comp_leb128} in Algorithm~\ref{alg:leb128}.

\begin{algorithm}[ht]
\caption{LEB128 algorithm}
\label{alg:leb128}
\scriptsize
\LinesNumbered 
\KwIn{input stream $inData$}
\KwOut{output stream $outData$}
\While{not reach the end of $inData$} {
\tcc{(s0) }
  $number \leftarrow$ read next  32-bit from $inData$ \label{algo:tdic32_read}\;
\tcc{(s1)}
  $zeros \leftarrow$ counting the leading zeros in $number$\;
  $encoded \leftarrow$ squeeze $zeros$ in $number$ under LEB128 format\;
  \tcc{(s2)}
    write  $encoded$ to $outData$ \label{algo:tdic32_write}\;
}
\end{algorithm}

\textbf{Stateful Compression.} 
In contrast, stateful compression in \design uses a maintained state to boost compressibility. While the stateful mode often provides superior compression ratios, it incurs additional overhead due to state maintenance. Algorithm~\ref{alg:stateful} illustrates the five-step procedure used in \design's stateful compression mode.

During \design's development, we explored three prevalent types of state usage in stream compression:

\begin{enumerate}
    \item \textbf{Value-based state} is the simplest state type, requiring only updates and records of the ``last compressed'' value. \design employs this state in scenarios where computational efficiency is of utmost importance. Techniques like delta encoding~\cite{comp_baseDelta,comp_tersecades} and run-length encoding (RLE)~\cite{wang2017experimental} that are considered ``lightweight'' for both stream and database compression~\cite{comp_tersecades, wang2017experimental}, serve as inspiration for this state type.
    
    \item \textbf{Dictionary-based state}, although requiring more memory, can greatly enhance compression ratios. A dictionary-based state maintains a collection of previously encountered ``last compressed'' values. \design leverages this type of state to optimize compression ratios when memory resources permit. It takes full advantage of the L1 cache~\cite{comp_lz4}, making dictionary-based compression particularly effective.
    
    \item \textbf{Model-based state} approximates data using a model with several parameters. This state type provides high compression ratios when the data stream closely aligns with a certain model. \design uses this state type for such data streams. The piece-wise linear approximation (PLA) technique has been particularly effective for compressing sensor-generated time series~\cite{zhou2011dissemination,blalock2018sprintz}.
\end{enumerate}

\begin{algorithm}
\caption{\design's Stateful Compression Mode}
\label{alg:stateful}
\LinesNumbered 
\scriptsize
\KwIn{input stream $inData$}
\KwOut{output stream $outData$}
\While{$inData$ is not stopped}
{
    (s0) load the tuple(s) from $inData$ \label{algo:stateful_S0} \;
    (s1) pre-process \label{algo:stateful_S1} \;
    (s2) state update  \label{algo:stateful_S2} \;
    (s3) state-based encoding\label{algo:stateful_S3}\;
    (s4) output compressed data to $outData$ \label{algo:stateful_S3}\;
}
\end{algorithm}

As a concrete example of stateful compression, we add the delta-encoding (i.e., value-based state) to Algorithm~\ref{alg:leb128}, and demonstrate the $Delta-LEB128$ in Algorithm~\ref{alg:delta-leb128}.

\begin{algorithm}[ht]
\caption{Delta-LEB128 algorithm}
\label{alg:delta-leb128}
\scriptsize
\LinesNumbered 
\KwIn{input stream $inData$}
\KwOut{output stream $outData$}
state $\leftarrow$ 0 \;
\While{not reach the end of $inData$} {
\tcc{(s0)}
  $number \leftarrow$ read next  32-bit from $inData$ \label{algo:delta_leb128_read}\;
\tcc{(s1)}
    lastNumber $\leftarrow$ state \;
\tcc{(s2)}
    state $\leftarrow$ number \;
\tcc{(s3)}
  $x$ $\leftarrow$ number-lastNumber \;
  $zeros \leftarrow$ counting the leading zeros in $x$\;
  $encoded \leftarrow$ squeeze $zeros$ in $x$ under LEB128 format\;
  \tcc{(s4)}
    write  $encoded$ to $outData$ \label{algo:delta_leb128_write}\;
}
\end{algorithm}

\subsubsection{Byte Alignment}
\label{subsub:byte_align}
The alignment strategy of a compression algorithm, specifically whether it is byte-aligned or not, greatly impacts both compressibility and computational overhead. \design offers both byte-aligned and non-byte-aligned modes to accommodate various computational environments and use cases.

\textbf{Byte-Aligned Compression.}
In byte-aligned mode, \design aligns data encoding with byte boundaries, a strategy similar to the $LEB128$ approach~\cite{comp_leb128}. This mode capitalizes on the hardware characteristics of modern processors that manage registers and memories in units of bytes. Despite this, a trade-off exists as the output code length must be an integer multiple of a byte (i.e., 8 bits), potentially leading to some waste of bits.

\textbf{Non-Byte-Aligned Compression.}
Non-byte-aligned mode in \design works to mitigate bit-level waste. However, this comes at the expense of additional computational overhead due to the need for more logical and bit-shift operations, such as AND/OR. Because the minimal unit of output code length is one bit instead of one byte, appending and extracting data operations may require extra computational instructions. Despite the increased overhead, this mode can significantly increase the compression ratio in specific scenarios, thereby providing a valuable option in the \design framework.

\begin{figure}[t]
    \centering
    \includegraphics*[width=0.3\textwidth]{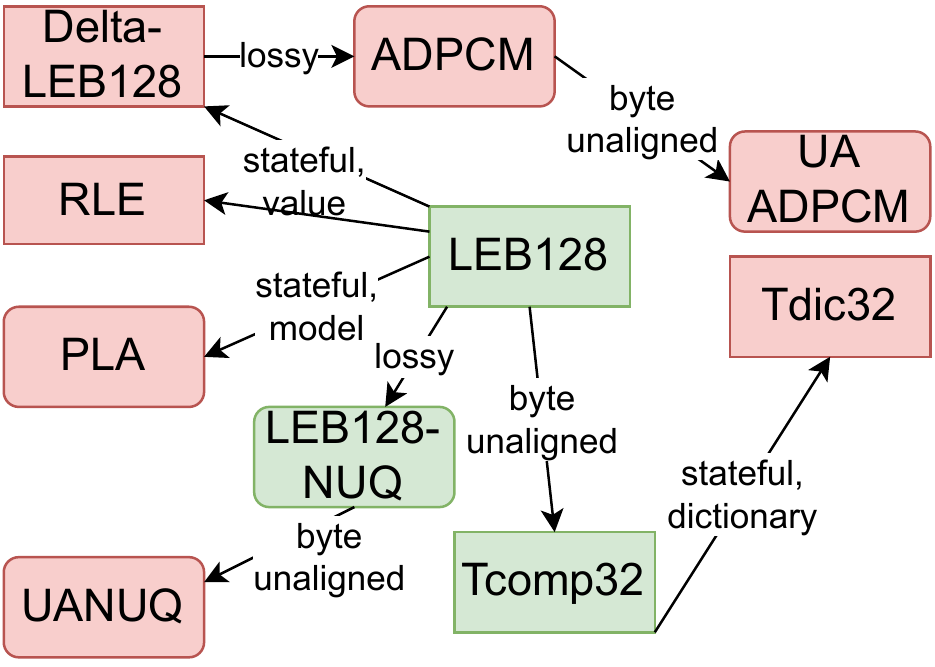}
    \caption{Incremental relationship among algorithms.}  \label{fig:algos_evo}
\end{figure}

\subsubsection{Compression Implementation}
\label{subsec:compression_implement}
The implementation of our algorithms follows a step-by-step refinement and optimization process, illustrated in Figure~\ref{fig:algos_evo}. This approach allows us to incrementally examine the impact of each algorithmic variation, comparing pairs of sequentially connected algorithms.

Initially, we implement the LEB128 algorithm in the same way as Android-Dex~\cite{comp_leb128}, which serves as our foundation for stateless and byte-aligned stream compression. Next, we adapt LEB128 to allow byte-unaligned output by simplifying Elias encoding~\cite{comp_elias} into $Tcomp32$. This adjustment involves removing leading zeros from each 32-bit tuple and calculating the prefix based on the length of non-zero bits.

By changing the encoding of the LEB128 into non-uniform quantization~\cite{comp_nonuniform}, we can implement the lossy compression $LEB128-NUQ$, which can be further equipped with byte-unaligned output by upgrading into $UANUQ$. Besides the stateless lossy extension, $LEB128$ and $Tcomp32$ can be also enriched with compression states.

$Tdic32$ utilizes a LZ4-like hash table~\cite{comp_lz4} as its state, which tracks duplication across 4096 table entries by default. Each 32-bit tuple is replaced by its shorter hash index if the tuple has appeared previously, as tracked by the table. In contrast, Delta-LEB128 maintains a simple state: it computes the difference between two consecutive values, termed as the "delta" state~\cite{comp_tersecades,comp_baseDelta}, and applies LEB128's method to output the "delta" state instead of the original value. The well-known commercial algorithm run-length-encoding ($RLE$)~\cite{wang2017experimental} resembles $Delta-LEB128$ in a value-based state, but it can further track the continuous duplication times of the state value.
Besides value and dictionary, we can also upgrade $LEB128$ into model-based stateful compression $PLA$~\cite{zhou2011dissemination,blalock2018sprintz}, which utilizes piecewise linear approximation as the model.

Finally, we implement a form of lossy compression by modifying the ``delta'' state encoding in Delta-LEB128, leading to the creation of the Adaptive Differential Pulse-Code Modulation (ADPCM). Specifically, we employ lossy non-uniform quantization~\cite{comp_nonuniform} in ADPCM to encode the ``delta'' state, instead of the lossless LEB128 style. The ADPCM can be further equipped with byte-unaligned, $Tcomp32$-like output format as $UAADPCM$. 
\begin{table}[]
\caption{Multicore edge processors studied.}
\label{tab:platforms}
\resizebox{\columnwidth}{!}{%
\begin{tabular}{|p{2cm}|p{1.7cm}|p{1cm}|p{1cm}|p{2cm}|p{1cm}|}
\hline
\textbf{Processor Name} & \textbf{Processor Architectures} & \textbf{Byte Length} & \textbf{ISA Pattern} & {Frequency Range/GHz} & {Core Number} \\ \hline
$H2+~\cite{amp_h2+}$        & SMP  & 32 bit & RISC & 0.416 $\sim$ 1.2 & 4 \\ \hline
$RK3399^{AMP}~\cite{amp_rk3399}$ & AMP  & 64 bit & RISC &  0.416 $\sim$ 1.8 & 6 \\ \hline
$RK3399^{SMP}~\cite{amp_rk3399}$ & SMP  & 64 bit & RISC & 0.416 $\sim$ 1.8 & 6 \\ \hline
$Z8350~\cite{amp_z8350}$      & SMP  & 64 bit & CISC & 0.8 $\sim$ 1.92 & 4 \\ \hline
\end{tabular}%
}
\end{table}
\subsection{Accommodating the Multicore Edge Landscape}
\label{sec:hardware_design}
This section discusses the fundamental aspects of the multicore edge hardware that influence \design's implementation, making it conducive to various edge environments. It delves into the choices of architecture and Instruction Set Architecture (ISA), and extends the discussion to the utilization and regulation of hardware resources. The edge processors under examination are detailed in Table~\ref{tab:platforms}.

\subsubsection{Processor Architecture: Symmetric and Asymmetric}
\label{subsec:arch_selection}
In the context of multicore processors, the architectural design can be broadly bifurcated into symmetric and asymmetric models.

\textbf{Symmetric Architecture.}
Symmetric multicore processors (SMPs) are not exclusive to center servers but are equally viable for edge devices. For instance, the UP board platform~\cite{amp_upBoard} supported by \design employs a 4-core SMP z8350~\cite{amp_z8350}. The primary distinction between the SMPs used at the edge and at the center lies in their energy efficiency: each core in an edge SMP is generally more energy-optimized and less powerful than its center counterpart. \design leverages the inherent simplicity and unified nature of the SMP architecture at the edge for straightforward parallel execution, facilitating the transfer of existing optimizations from cloud SMPs~\cite{comp_para_hsi,comp_para_bzip2}.

\textbf{Asymmetric Architecture.}
Aiming for a balanced trade-off between performance and energy efficiency, asymmetric architectures offer a compelling alternative. By featuring various types of execution units on a single processor, these architectures open up new avenues for optimization in stream compression. This work emphasizes asymmetric architectures within single ISA, known as asymmetric multicore processors (AMPs)~\cite{amp_survey_2016}. However, it acknowledges the potential of other forms of asymmetry, such as edge CPU+GPU~\cite{edge_cpu_gpu,edge_cpu_gpu_2}, CPU+DSP~\cite{couple_cpu_dsp}, and CPU+FPGA~\cite{yang2015adaptive}, as promising areas for future exploration.

\subsubsection{Instruction Set Architecture}
\label{subsec:isa_selection}
\design's flexibility shines through its support for a variety of ISAs, thereby covering the broad scope of edge devices.

\textbf{32-bit vs. 64-bit.}
While 32-bit processors maintain a foothold in the edge domain due to their long-standing development ecosystem and lower cost, 64-bit processors are increasingly gaining traction due to their enhanced memory addressing efficiency and speed. \design extends its support to both variants, facilitating a balanced comparison within the realm of stream compression.

\textbf{CISC vs. RISC.}
The trade-off between the flexibility of complex (CISC) and the efficiency of reduced (RISC) instruction set architectures is dependent on specific applications. \design explores this dynamic in the context of stream compression. It recognizes the growing popularity of RISC architectures, especially ARM processors, for edge devices~\cite{edge_survey}.

\subsubsection{Energy-Efficient Resource Management}
\label{subsec:resource_regulation}
Once a specific architecture and ISA have been selected, \design provides the means to finely tune hardware resources, thereby enhancing the balance between processing time and energy consumption.

\textbf{Frequency Regulation.} The processor's clock frequency directly influences its performance and power consumption. By executing more instructions per unit of time, a higher frequency leads to reduced processing time and increased throughput but at the cost of elevated energy consumption. Conversely, lower frequencies are more energy-efficient but result in lower processing rates. \design facilitates exploration of this trade-off by allowing stream compression tasks to be run at varying frequency settings. Moreover, the capability extends to dynamic frequency scaling using Dynamic Voltage and Frequency Scaling (DVFS) technology~\cite{amp_DVFS1,amp_DVFS2,amp_DVFS3}. Though changes in frequency introduce overheads, \design offers the potential to explore whether the advantages of DVFS outweigh these costs in the context of stream compression.

\textbf{Core Number Regulation.} While cloud-based stream processing frameworks~\cite{sp_briskstream,sp_coprocessing} typically utilize all available computational resources in pursuit of maximum performance, edge-based systems need to be more energy-conscious. Consequently, it might be beneficial to turn off or idle certain processor cores during stream compression operations to conserve energy. \design enables investigation into this trade-off between energy consumption and processing throughput by allowing variable core usage, thereby further enhancing its adaptability to diverse edge environments.
\subsection{Integrated Software and Hardware Optimization}
\label{subsec:integration}
\design optimizes stream compression by finely tuning software strategies and hardware configurations to create a synergistic interplay that maximizes both efficiency and performance. Central to this integration is \design's comprehensive suite of parallelization strategies. The nuances of this integration will be unpacked in the subsequent sections.

\subsection{Execution Strategies}
The choice between Eager and Lazy Execution depends on stream data characteristics and hardware resources. 

\textbf{Eager Execution.} 
Eager execution aligns closely with the inherent nature of streaming data—infinitely incremental~\cite{comp_ASE,comp_stream_dcc2019}. As each tuple arrives, it is compressed instantly. However, while this strategy reflects the streaming model, it can lead to inefficient hardware utilization due to frequent partitioning, synchronization, and potential cache thrashing.

\textbf{Lazy Execution.} 
To counteract the potential inefficiencies of Eager Execution, \design introduces Lazy Execution. This strategy batches several tuples together before compression, a practice known as \emph{micro-batching}~\cite{sp_spark,comp_tersecades}. While this approach reduces communication overhead and promotes better hardware utilization, selecting an appropriate batch size can be challenging, necessitating a careful balance between frequent cache flush and reload operations and not overburdening slower memory storage.

\subsubsection{State Management Strategies}
Parallelizing stateful stream compression demands careful consideration of state management and sharing. \design provides three key strategies: State Sharing and Private State.

\textbf{State Sharing.} State Sharing allows all threads to share a single state, with locks implemented to prevent write conflicts. While this method can theoretically offer higher compressibility due to a collective record of past tuples, concurrency control overhead may hinder parallelism and impact performance.

\textbf{Private State.} In contrast, a private state lets each thread maintain its own state, thereby eliminating concurrent conflicts. This strategy maximizes parallelism but potentially reduces compressibility, as the isolated states limit a thread's awareness of tuples handled by others.


\subsubsection{Scheduling Strategies}
\label{subsubsec:scheduling}
For efficient workload distribution across multicore processors, \design utilizes two main scheduling strategies: Simple Uniform Scheduling and Asymmetric-aware Scheduling.

\textbf{Simple Uniform Scheduling.} In Symmetric Multicore Processor (SMP) environments, \design uses a ``balanced partition and equal distribution ratio'' approach to scheduling~\cite{comp_para_bzip2}. This method takes advantage of the symmetry inherent in SMPs, where all cores have equal computational capacity and communication distance.

\textbf{Asymmetric-aware Scheduling.} For Asymmetric Multicore Processors (AMPs), which have cores with different computational abilities and communication distances, \design employs Asymmetric-aware Scheduling, which is described in detail in our preliminary work~\cite{CStreamICDE}. This strategy tailors workload distribution to the computational intensity of different stream compression stages, thereby optimizing hardware utilization.
\begin{figure}[t]
    \centering
    \includegraphics*[width=0.32\textwidth]{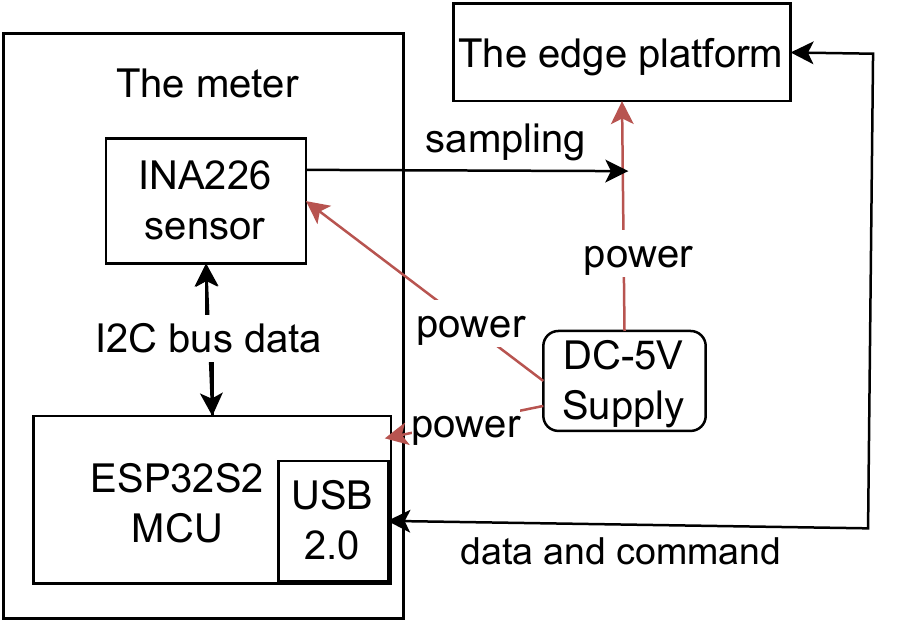}
    \caption{The design of energy meter.}  \label{fig:meter}   
\end{figure}
\subsection{Energy Metering}
\label{subsec:energy_metering}
Understanding and accurately measuring energy consumption is of paramount importance in edge computing, where power constraints are often more stringent than in conventional data centers. To support this need, \design includes a custom-developed energy meter, as illustrated in Figure~\ref{fig:meter}, to accurately measure energy consumption across various edge platforms.
The detailed specification is shown in Table~\ref{tab:meter}.

This energy metering component consists of Texas Instrument's INA226~\cite{amp_ina226} chip, which acts as a sensor for current and voltage, and the Espressif's ESP32S2~\cite{amp_esp32s2} micro control unit (MCU) for data pre-processing and USB-2.0 communication with targeted asymmetric multicores. In this configuration, the meter offers precise measurement capabilities while maintaining a low overhead. Additionally, for the UP board platform, it is possible to directly read energy consumption data from X64 msr registers~\cite{amp_msr} before and after an experimental run. 

\textbf{Accurate Measurement.} The integration of the INA226 chip and the ESP32S2 MCU in our energy metering system facilitates high-precision measurement of energy usage. This accuracy extends from the overall system level down to the granularity of individual operations, providing deep insights into the energy profile of the stream compression process.

\textbf{Low Overhead.} Our energy metering system has been designed for minimal overhead. The ESP32S2 MCU handles data pre-processing, and the use of USB-2.0 communication ensures that the monitoring process does not interfere with the primary tasks of stream compression and parallel execution.

\textbf{Real-time Feedback.} The energy metering system offers real-time feedback on energy consumption. This feedback enables dynamic adjustments of execution strategies, state management, and scheduling based on current energy usage, thus supporting the goal of energy-efficient stream compression.

\textbf{Comprehensive Integration.} The energy metering system is tightly integrated with the rest of the \design framework. This integration allows it to work synergistically with \design's software strategies and hardware configurations, guiding optimizations and adaptations based on up-to-date energy consumption data.


\begin{table}
\caption{Specifications of the energy meter
}
\centering
\label{tab:meter}
\resizebox{.75\columnwidth}{!}{%
\begin{tabular}{|L{2cm}|L{5cm}|L{0cm}}
\hline
Sensor chip       & INA226       \\ \hline
MCU chip          & ESP32S2      \\ \hline
Interface         & USB 2.0 Full \\ \hline
Raw resolution    & 16 bits      \\ \hline
LSB               & 1.25 mV, 2mA \\ \hline
Measurement range & 0-15V, 0-4A  \\ \hline
Sample rate       & 1K SPS       \\ \hline
\end{tabular}%
}
\end{table}

\section{Methodology}
\label{sec:methodology}
This section presents our methodological approach to assessing the effectiveness and performance of CStream. It provides a comprehensive explanation of the chosen performance metrics, the benchmark workloads utilized, and the selection and characteristics of the hardware platforms involved in the evaluation.

\subsection{Performance Metrics}
\label{subsec:performance_metrics}
In our evaluation of \design's performance, we focus on a range of metrics, each providing insight into different aspects of its functionality. These performance indicators, initially introduced in Section~\ref{sec:preliminaries}, include \emph{compression ratio}, \emph{throughput}, \emph{energy consumption}, and \emph{end-to-end latency}. For each metric, we aim to establish a consistent measurement approach. The average value of these metrics is calculated over a substantial data volume to avoid fluctuations and ensure measurement consistency - specifically, over 932800 bytes of tuples.

The \emph{compression ratio} and \emph{normalized root mean square error (NRMSE)} are calculated by comparing the compressed data against the raw input. The NRMSE, specifically used for lossy compression, is defined as $NRMSE=\frac{1}{\bar x}\times \sqrt{\frac{\sum_i^N (x[i]-y[i])^2} {N}}$, where $\bar x$ represents the average value of the input data stream, $x[i],y[i]$ denote the individual values of the raw input and the reconstructed data from compression, respectively, and $N$ is the total input data volume. 

\emph{Throughput} and \emph{end-to-end latency} provide insights into the system's operational efficiency. Throughput is measured as $throughput=\frac{N}{processing \ time}$, where the processing time is obtained using OS-specific APIs such as \emph{gettimeofday}. End-to-end latency, an important metric in edge computing scenarios requiring real-time or near-real-time data processing, quantifies the total time for a data element to traverse the compression system from input to output. 

\emph{Energy consumption} evaluation follows a two-step procedure. First, we measure and eliminate the static energy consumption caused by irrelevant hardware or software components, such as the Ethernet chip and back-end tcp/ip threads. Next, we monitor the energy consumption during the running of the stream compression benchmark, without incurring additional overhead or interference. The specifics of energy recording are platform-dependent and are discussed in Section~\ref{subsec:energy_metering}. 

\subsection{Benchmark Workloads}
\label{subsec:workload}
In order to comprehensively evaluate \design across a diverse range of IoT use cases (discussed in Section~\ref{subsec:sc_at_edge}), we use five representative IoT datasets. These datasets were chosen to reflect various data sources (single and multiple) and diverse data structures (plain, binary structured, and textual structured). Furthermore, we assess the compressibility of data from two perspectives: \emph{stateless compressibility} and \emph{statefu; compressibility}, utilizing a synthetic dataset to calibrate these properties.

The \emph{stateless compressibility} refers to the compressible space within each individual tuple $x_t=v$, which can be exploited by both stateless and stateful stream compression (refer to Section~\ref{subsec:intro_state}). On the other hand, \emph{stateful compressibility} points to the compressible space hidden in the context of the data stream, considering the current tuple $x_{\tau}$ and some past tuples $\{x_t|t<\tau\}$ together. This can only be fully exploited by a suitable stateful compression.

Our selected datasets are summarized in Table~\ref{tab:datasets} and are detailed below:

\begin{enumerate}
    \item \textbf{ECG}~\cite{comp_mit_ecg}: The ECG dataset consists of raw ADC recordings from electrocardiogram (ECG) monitoring provided by the MIT-BIH database. Each reading is packaged as a plain 32-bit value for our evaluation. As ECG is a direct, unstructured reflection of a continuous physical process, it exhibits the highest levels of both independent and associated compressibility.
    \item \textbf{Rovio}~\cite{sp_rovio}: The Rovio dataset continuously monitors user interactions with a specific game to ensure optimal service performance. Each data entry consists of a 64-bit key and 64-bit payload. The Rovio dataset exhibits two compressible traits: first, its payload is constrained to a relatively small dynamic range, indicating independent compressibility; second, different tuples may share the same key, which demonstrates associated compressibility.
    \item \textbf{Sensor}~\cite{ds_sensor}: The Sensor dataset is comprised of full-text streaming data generated by various automated sensors (e.g., temperature and wind speed sensors). For our evaluation, every 16 ASCII characters in the Sensor dataset forms one 128-bit tuple. The Sensor dataset primarily exhibits associated compressibility due to the repetition of several fixed XML patterns across different tuples.
    \item \textbf{Stock}~\cite{ds_stock} and \textbf{Stock-Key}~\cite{ds_stock}: The Stock dataset is a real-world stock exchange dataset packed in a (32-bit key, 32-bit payload) binary format. It exhibits less compressibility than Rovio due to fewer key duplications. Stock-Key is a subset of the Stock dataset, containing only the 32-bit keys.
    \item \textbf{Micro}: The Micro dataset is a synthetic 32-bit dataset~\cite{sp_intra_exp}, used to further tune independent and associated compressibility. We can adjust its dynamic range to control independent compressibility and its level of duplication to control associated compressibility.
\end{enumerate}

To mitigate the impact of network transmission overhead, all input datasets are preloaded into memory before testing. Each tuple is assigned a timestamp (starting from 0) to reflect its actual arrival time to the system, and tuples are time-ordered. These timestamps, which are stored separately from each tuple and are not subjected to compression, help provide a realistic simulation of data arrival in a real-world scenario. Unless otherwise specified, we generate incremental timestamps evenly to simulate an average arrival speed of $16\times10^6$ bytes per second (e.g., $10^6$ tuples per second for the Rovio dataset, which consists of 128-bit tuples).

\subsection{Edge Computing Platforms} 
\label{subsec:implentation_meter}
To ensure a comprehensive and hardware-variant evaluation, we deploy three distinct edge computing platforms, each featuring unique characteristics as outlined in Table~\ref{tab:boards}. Importantly, all platforms are compatible with the mainline Linux kernel and support the Glibc with C++20 features, thereby enabling a shared codebase.

The Rockpi 4a~\cite{amp_rockpi4} serves as our default evaluation platform. Unless otherwise specified, we use its processor as $RK3399^{AMP}$ and engage each core to operate at its maximum frequency: 1.8GHz for the larger cores (core4 to core5) and 1.416GHz for the smaller cores (core0 to core3).

\begin{table}[]
\caption{Dataset Studied}
\label{tab:datasets}
\resizebox{\columnwidth}{!}{%

\begin{tabular}{|p{1.8cm}|p{1.2cm}|p{1.5cm}|p{2cm}|p{2cm}|}
\hline
Dataset Name & Source &Data Structure & Stateless Compressibility & Stateful Compressibility \\ \hline
ECG~\cite{comp_mit_ecg}          & single & plain         & high                    & high                     \\ \hline
Rovio~\cite{sp_rovio}        &multiple& binary structured  & medium                    & medium                   \\ \hline
Sensor~\cite{ds_sensor}       &multiple& textual structured   & low                      & high                     \\ \hline
Stock~\cite{ds_stock}        &multiple& binary structured    & low                       & medium                   \\ \hline
Stock-key~\cite{ds_stock}    &multiple & plain      & low                       & medium                   \\ \hline
Micro~\cite{sp_intra_exp}    &single  & plain & adjustable & adjustable \\ \hline
\end{tabular}%
}
\end{table}
\begin{table}[t]
\centering
\caption{Evaluated hardware platforms}
\label{tab:boards}
\resizebox{0.8\columnwidth}{!}{%
\begin{tabular}{|p{2cm}|p{1.5cm}|p{1.5cm}|p{1.5cm}|}
\hline
Model& Supported processor &Number of cores & Installed memory   \\ \hline
Banana pi zero-m2~\cite{amp_bananapi}    & $H2+$  &4    & 512MB       \\ \hline
Rockpi 4a~\cite{amp_rockpi4}       & $RK3399$   &2 big + 4 little     & 2GB    \\ \hline
UP board~\cite{amp_upBoard}    & $Z8350$  &4    & 1GB    \\ \hline

\end{tabular}%
}
\end{table}

\section{Evaluation} 
\label{sec:evaluation}
In this section, we embark on an experimental journey to evaluate the potency of various stream compression schemes, particularly on edge platforms. This evaluation focuses on a strategic software-hardware co-design as explicated in Section~\ref{sec:co-design}. We delve into five key areas: 

\begin{enumerate}
\item An end-to-end case study of \design's solution space is demonstrated in Section~\ref{subsec:e2eCase}
\item The selection and performance of different stream compression algorithms, are explored in Section~\ref{subsec:eva_algo}.
\item The impact and considerations of hardware variants, are investigated in Section~\ref{subsec:evalu_hard}.
\item The effectiveness and scalability of novel parallelization strategies, are examined in Section~\ref{subsec:evalu_para}.
\item The sensitivity and adaptability to varying workload characteristics, assessed in Section~\ref{subsec:evalu_workload}.
\end{enumerate}

Our exploration culminates in Section~\ref{subsec:summary}, where we collate our findings. Unless otherwise specified, we adopt the following parallelization strategies: 1) lazy execution with a 400-byte micro batch, 2) a regulated multicore workload distribution ratio to optimize throughput, and 3) the use of a non-shared state for stateful stream compression.
\subsection{End to End Case Study}
\label{subsec:e2eCase}

\begin{figure}[t]
    \centering
    \includegraphics*[width=0.85\columnwidth]{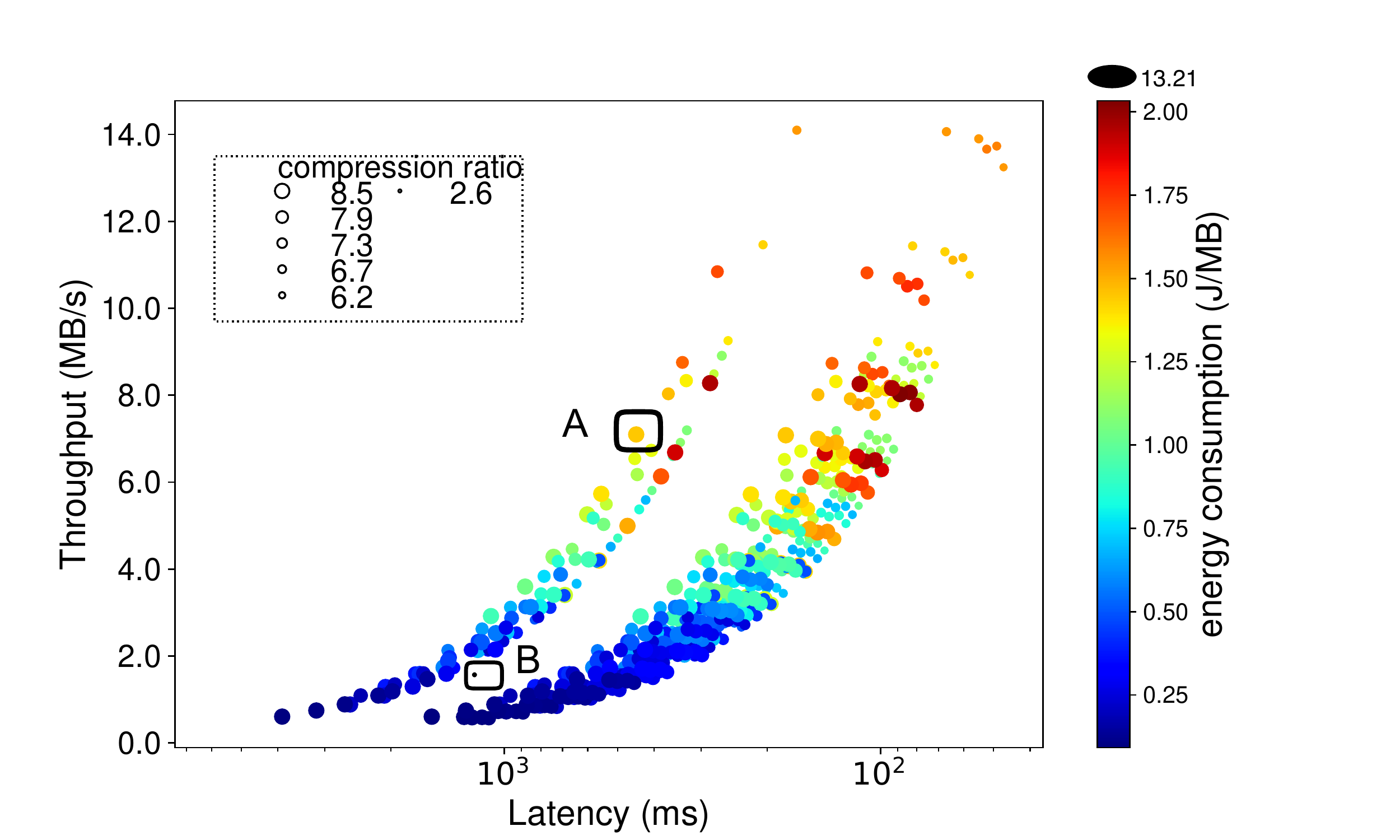}
    \caption{The solution space of \design under $\geq 6.0$ compression ratio, $\leq 5\%$ NRMSE, and RK3399 hardware.} \label{fig:impact_best_worst}   
\end{figure}

In this case study, we have ECG data to be compressed on RK3399 hardware, the compression ratio is expected to be larger than $6.0$, while the NRMSE should be controlled below $5\%$. \design will hence choose the $PLA$ algorithm, and the whole 
available solution is provided as colorful points in Fig~\ref{fig:impact_best_worst}.
In general, higher energy consumption is required when higher throughput, higher compression ratio, or lower latency is expected, and the specific optimal solution depends on users' prioritization of performance metrics.
For instance, if the user further wants to maximize the compression ratio, and then maximize throughput, while keeping the energy consumption within $1.5J/MB$, the optimal one is labelled as point A. Specifically, it utilizes 1 big core and 1 little core under their highest frequency, lets each of them use a private PLA state, and schedules the workload in an Asymmetric-aware manner. Both cores execute the stream compression under a $8KB$ micro-batch.

We also marked a careless solution point B in Fig~\ref{fig:impact_best_worst} as a contrast. This solution conducts a $Tdic32$ compression on 2 big cores and 4 little cores, the $Tdic32$ state is shared by all cores, and OS scheduling is used along with \emph{on-demand} DVFS. All cores use an eager strategy in conducting stream compression. Note that, the optimal solution A achieves $2.8 \times$ compression ratio, $4.3 \times$ throughput, $65\%$ latency reduction, and $89\%$ energy consumption reduction simultaneously than the careless solution B.

\begin{figure}
\begin{center}
	{\setlength{\fboxsep}{1pt}
			\hspace{0.99em}
				\begin{minipage}{0.45\textwidth}
					\includegraphics[width=\columnwidth]
					{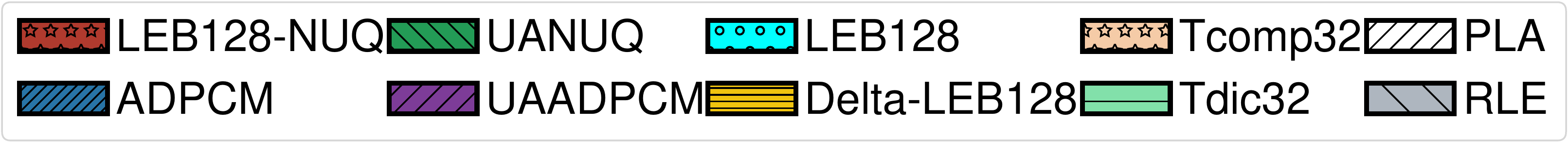}
				\end{minipage}
	}
        \centering
        \begin{subfigure}[t]{0.5\columnwidth}
            \centering
            \includegraphics[width=\columnwidth]{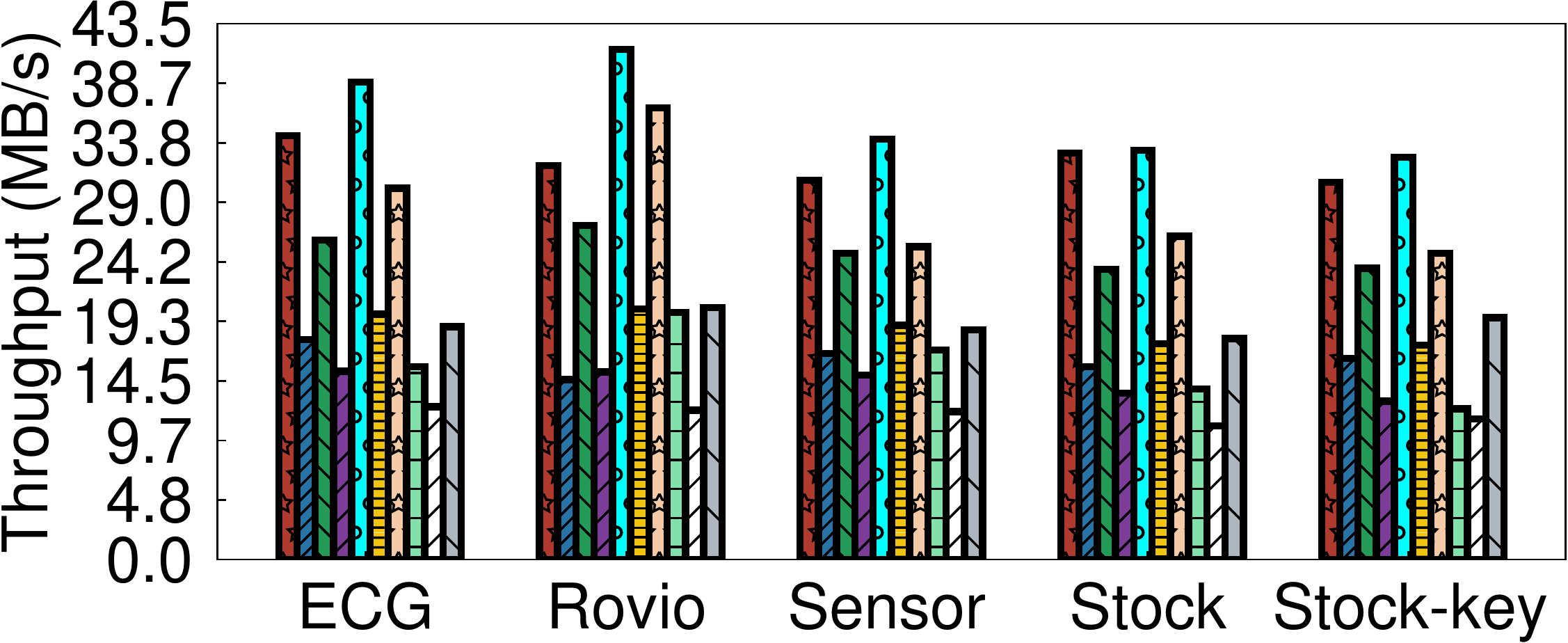}
            \caption{Throughput}
            \label{fig:thr_each}
        \end{subfigure}%
        ~ 
        \begin{subfigure}[t]{0.5\columnwidth}
            \centering
            \includegraphics[width=\columnwidth]{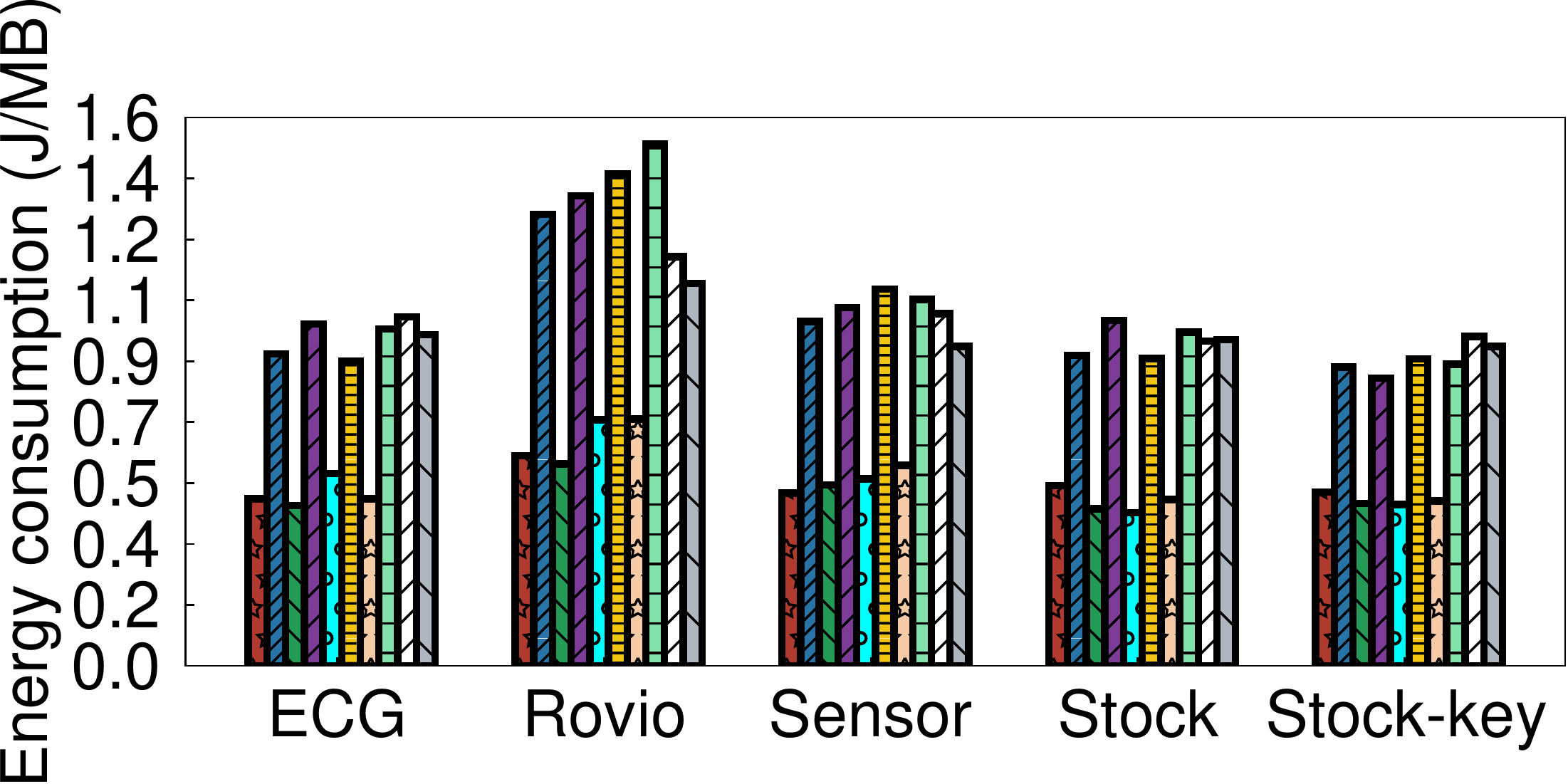}
            \caption{Energy consumption}
            \label{fig:ene_each} 
        \end{subfigure}
        \vspace{+5pt}
        
        \begin{subfigure}[t]{0.5\columnwidth}
            \centering
            \includegraphics[width=\columnwidth]{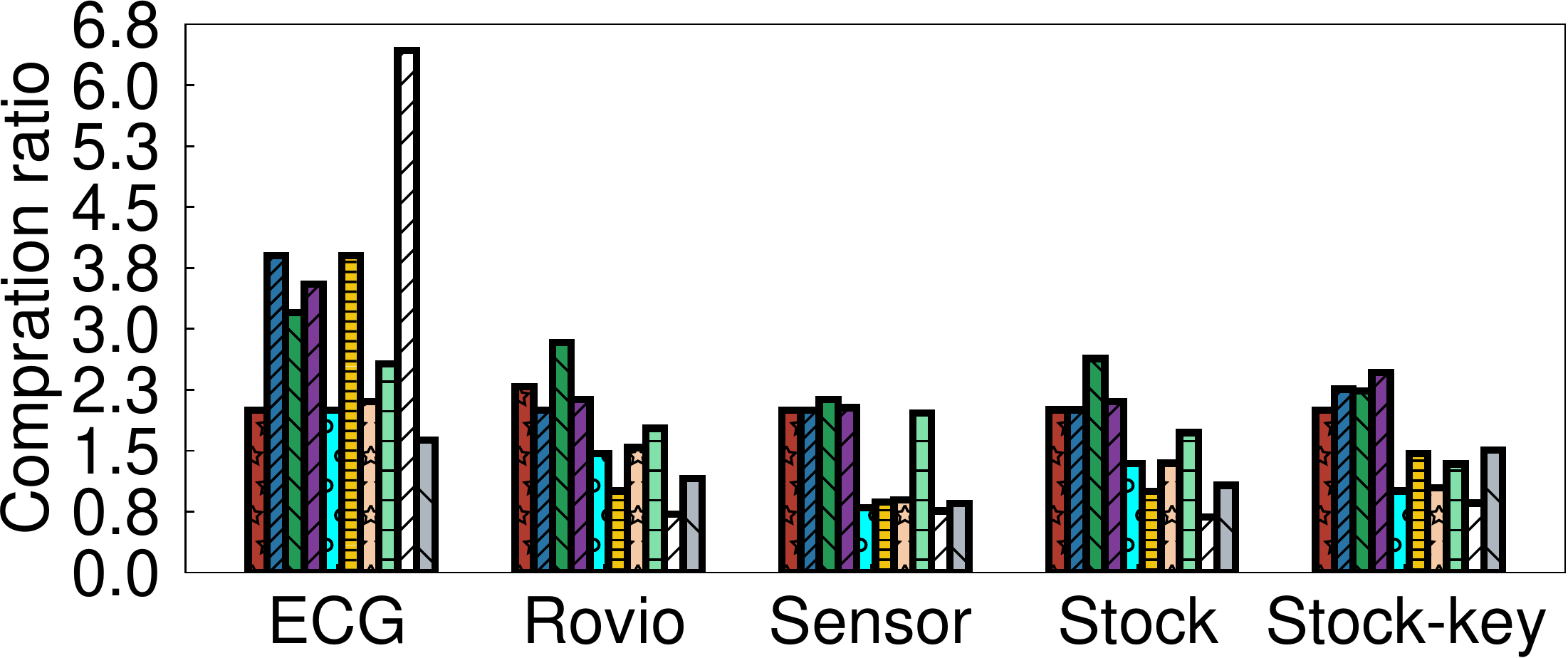}
            \caption{Compression ratio}
            \label{fig:cr_each}
        \end{subfigure}%
        ~ 
        \begin{subfigure}[t]{0.5\columnwidth}
            \centering
            \includegraphics[width=\columnwidth]{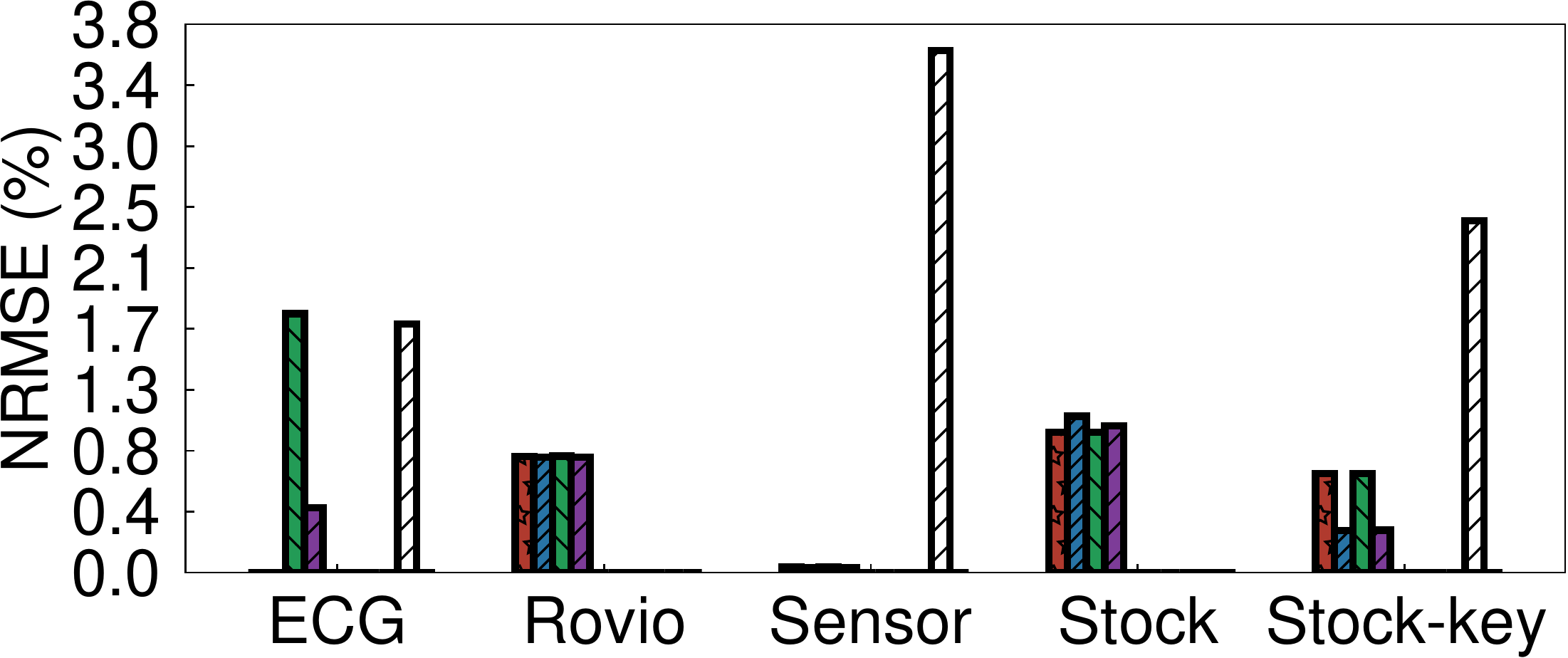}
            \caption{NRMSE}
            \label{fig:nrmse_each} 
        \end{subfigure}
        \caption{Comparing ten algorithms on five datasets.}        
\end{center}    
\end{figure}

\subsection{Algorithm Evaluation}
\label{subsec:eva_algo}
In order to evaluate the performance of \design's support for diverse stream compression algorithms, we test ten distinct algorithms, as summarized in Table~\ref{tab:algo_list}, on five real-world IoT datasets, as detailed in Table~\ref{tab:datasets}. 

\subsubsection{Fidelity: Lossless vs. Lossy Compression}

Referencing Figures~\ref{fig:thr_each}, ~\ref{fig:ene_each}, ~\ref{fig:cr_each}, and ~\ref{fig:nrmse_each}, the distinction between lossless and lossy compression becomes apparent. Lossy compression, represented by $LEB128-NUQ$, offers a superior compression ratio (between $2.0$ and $6.0$) across all datasets. Conversely, $LEB128$, a lossless algorithm, struggles to exceed a ratio of $2.0$. Remarkably, this high compression ratio by $LEB128-NUQ$ introduces only marginal information loss, with a NRMSE below $3.8\%$. 

\subsubsection{State Utilization: Stateless vs. Stateful Compression}

On comparing $Tcomp32$ (stateless) with $Tdic32$ (stateful, dictionary-based), we observe a trade-off between compression cost and effectiveness. While $Tcomp32$ offers more modest compression ratios, it presents a less complex, lower-cost compression process. On the other hand, $Tdic32$, despite its higher processing cost due to dictionary-based state management, excels in handling text data streams with high associated but low independent compressibility, like the Sensor dataset. 

\subsubsection{State Implementation: Value, Dictionary, and Model}

Further comparisons among $ADPCM$ (stateful, value-based), $Tdic32$ (stateful, dictionary-based), and $PLA$ (stateful, model-based) illustrate how different state implementations can impact compression performance. While $ADPCM$ consistently leads in throughput and energy consumption, $Tdic32$ and $PLA$ shine in handling structured data or data from multiple sources, offering higher compression ratios. 

\subsubsection{Byte Alignment Variations}

When comparing byte-aligned $LEB128$ and byte-unaligned $Tcomp32$, we note a trade-off between compression ratio and computational cost. $Tcomp32$ achieves higher compression ratios but at the expense of added computational overhead, given its bit-by-bit encoding process.

\begin{figure}[t!]
    \centering
    \begin{subfigure}[t]{0.5\columnwidth}
        \centering
        \includegraphics[width=\columnwidth]{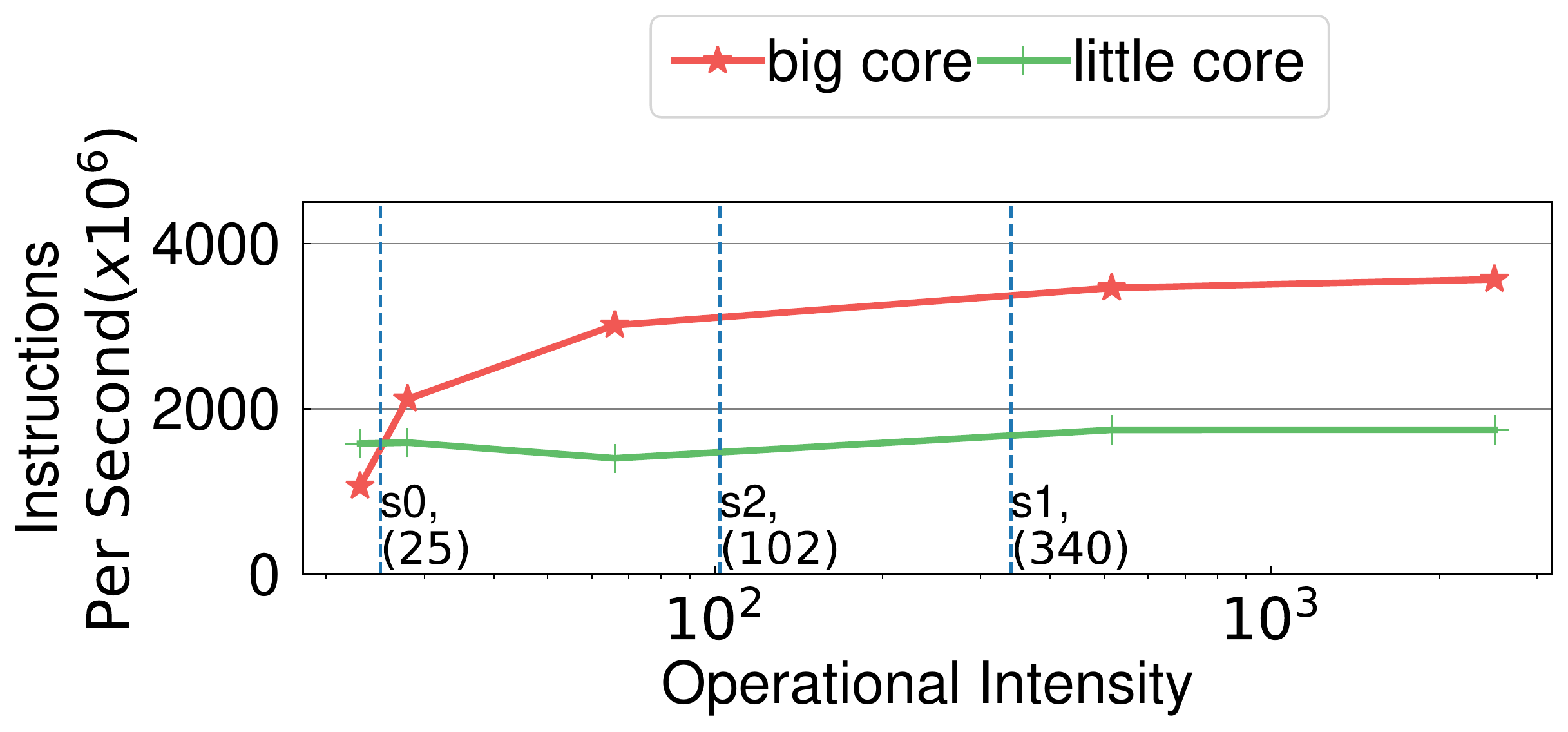}
        \caption{The roofline model on RK3399 processor}
        \label{fig:roofline}
    \end{subfigure}%
    ~ 
    \begin{subfigure}[t]{0.5\columnwidth}
        \centering
        \includegraphics[width=\columnwidth]{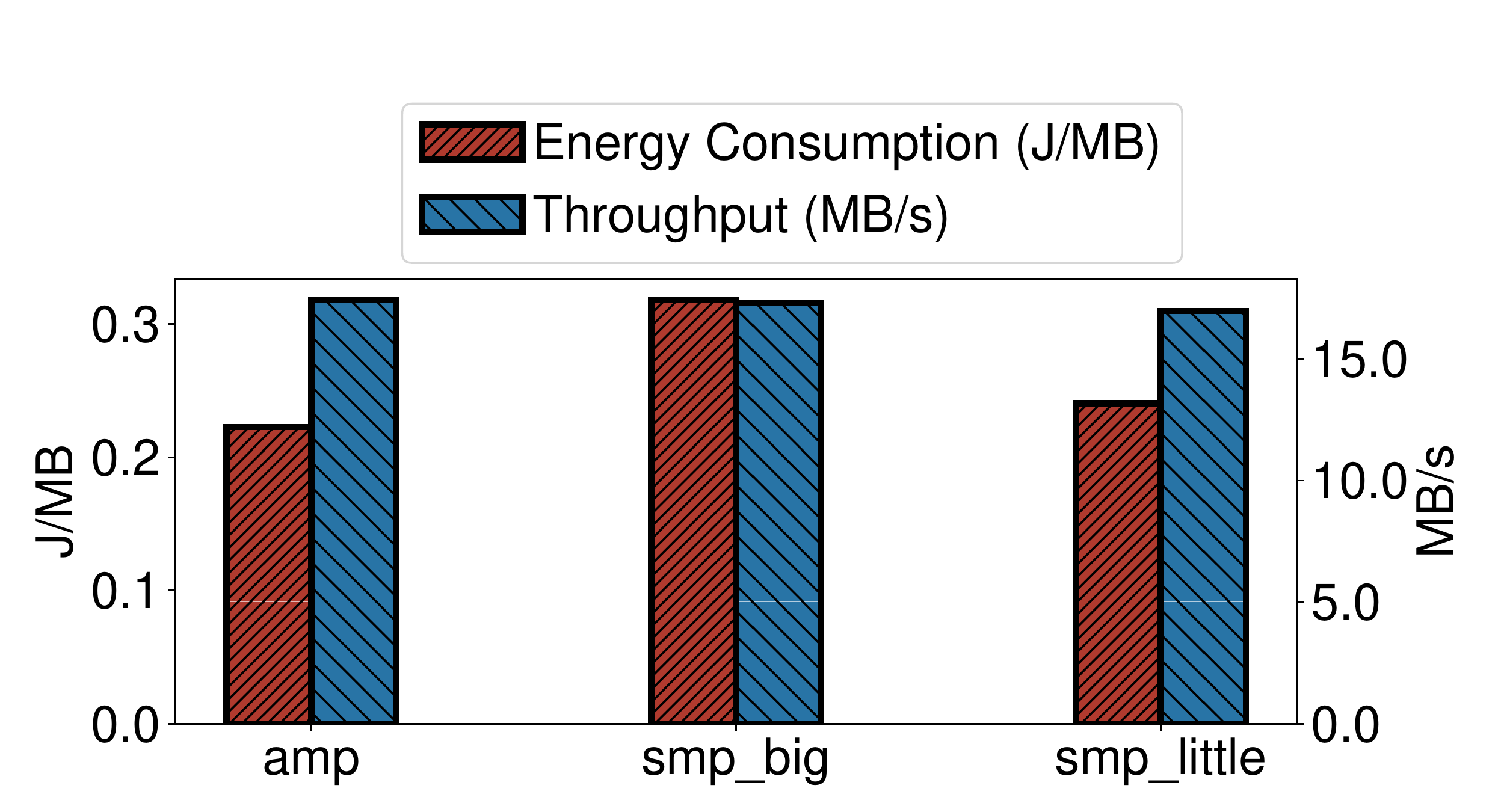}
        \caption{Energy consumption and throughput comparison}
        \label{fig:smp_amp} 
    \end{subfigure}
    \caption{Impacts of processor architectures for $Tcomp32$.}
\end{figure}

\begin{figure}[t!]
    \centering
    \begin{subfigure}[t]{0.45\columnwidth}
        \centering
        \includegraphics[width=\columnwidth]{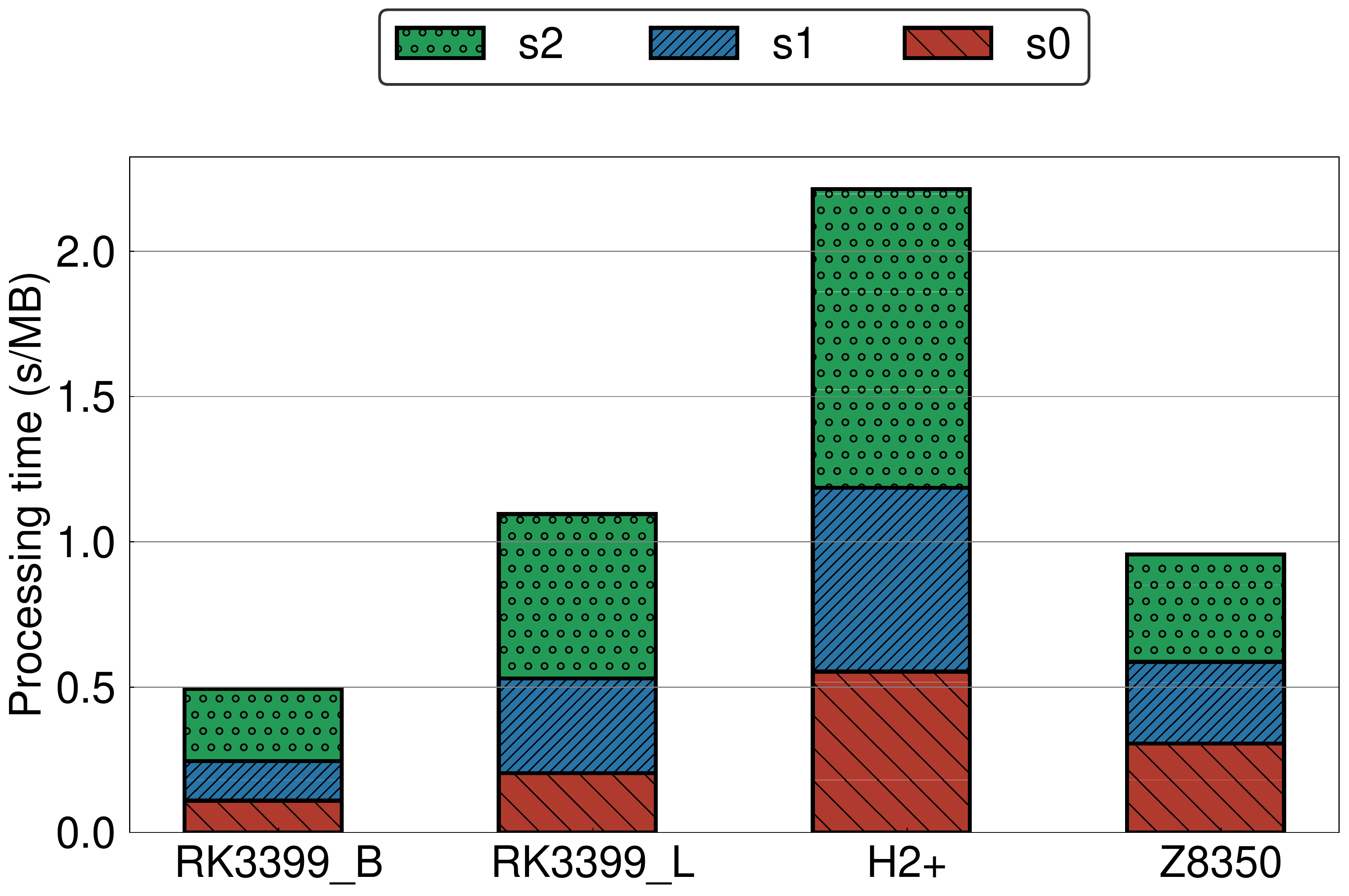}
        \caption{Unit time cost}
        \label{fig:unitTime}
    \end{subfigure}%
    ~ 
    \begin{subfigure}[t]{0.45\columnwidth}
        \centering
        \includegraphics[width=\columnwidth]{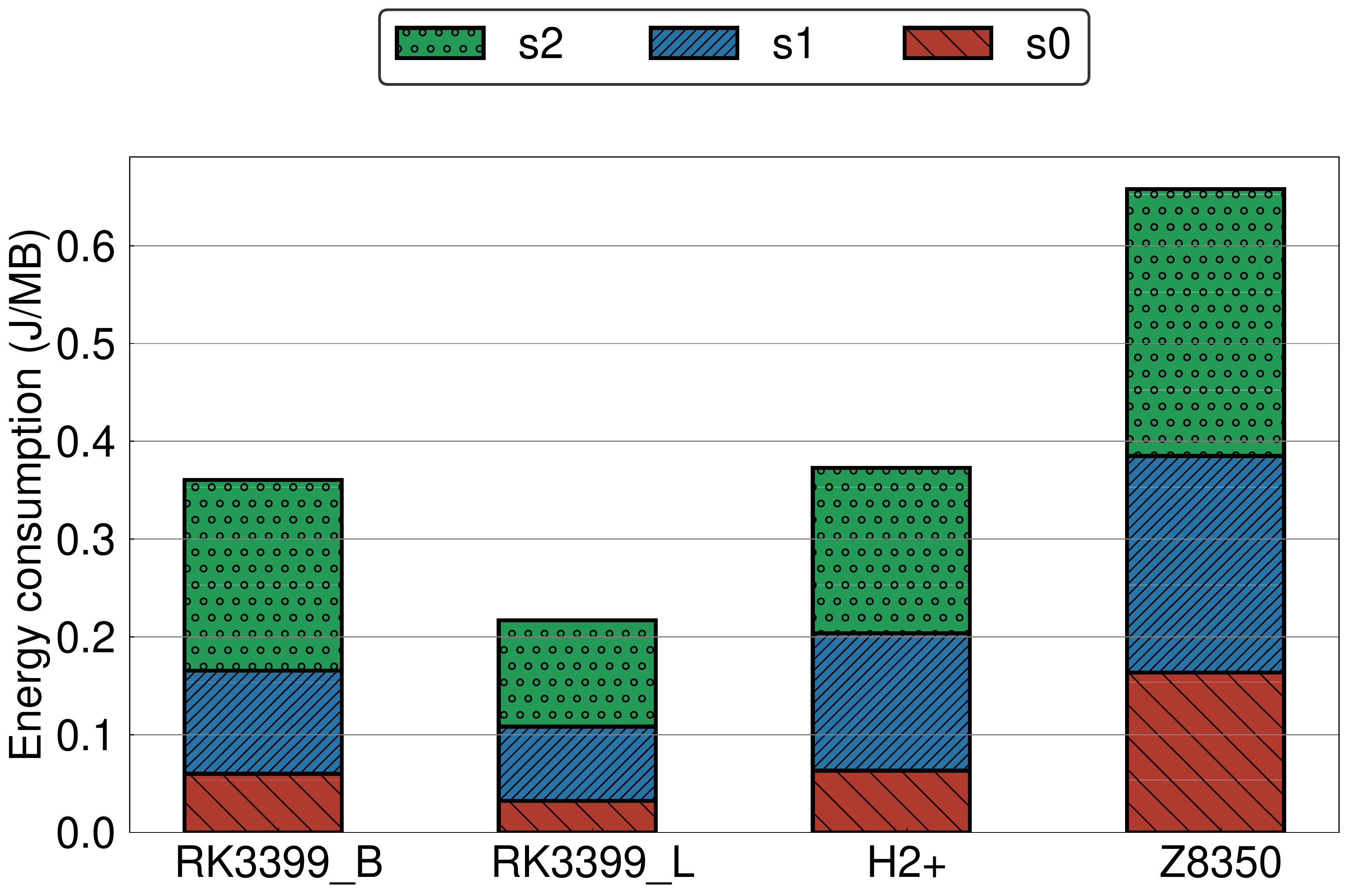}
        \caption{Unit energy cost}
        \label{fig:unitEnergy} 
    \end{subfigure}
    \caption{Impacts of ISA for $Tcomp32$.}
\end{figure}

\begin{figure}[t!]
    \centering
    \begin{subfigure}[t]{0.5\columnwidth}
        \centering
        \includegraphics[width=\columnwidth]{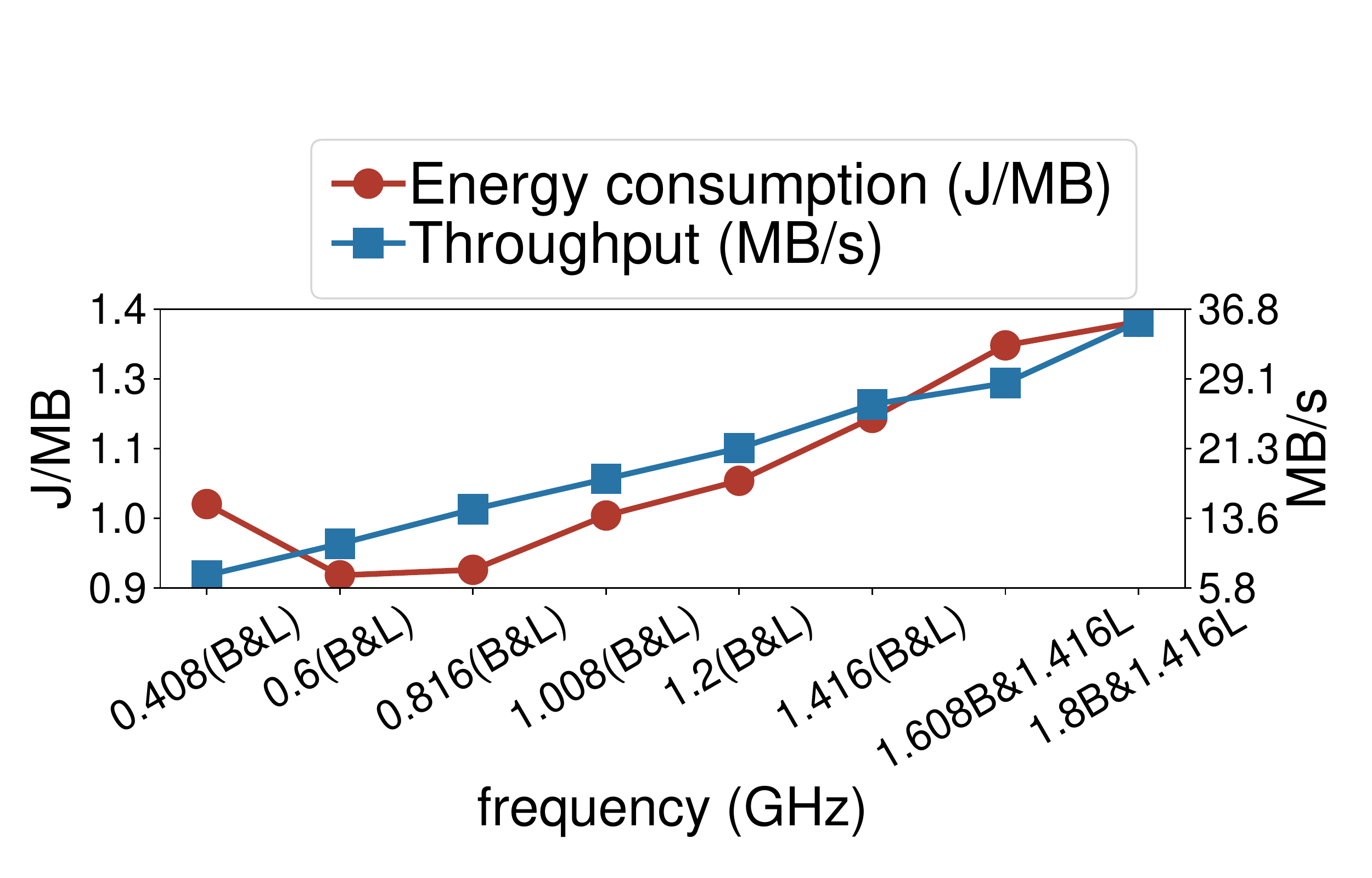}
        \caption{Statically tuning}
        \label{fig:freq_ene}
    \end{subfigure}%
    ~ 
    \begin{subfigure}[t]{0.5\columnwidth}
        \centering
        \includegraphics[width=\columnwidth]{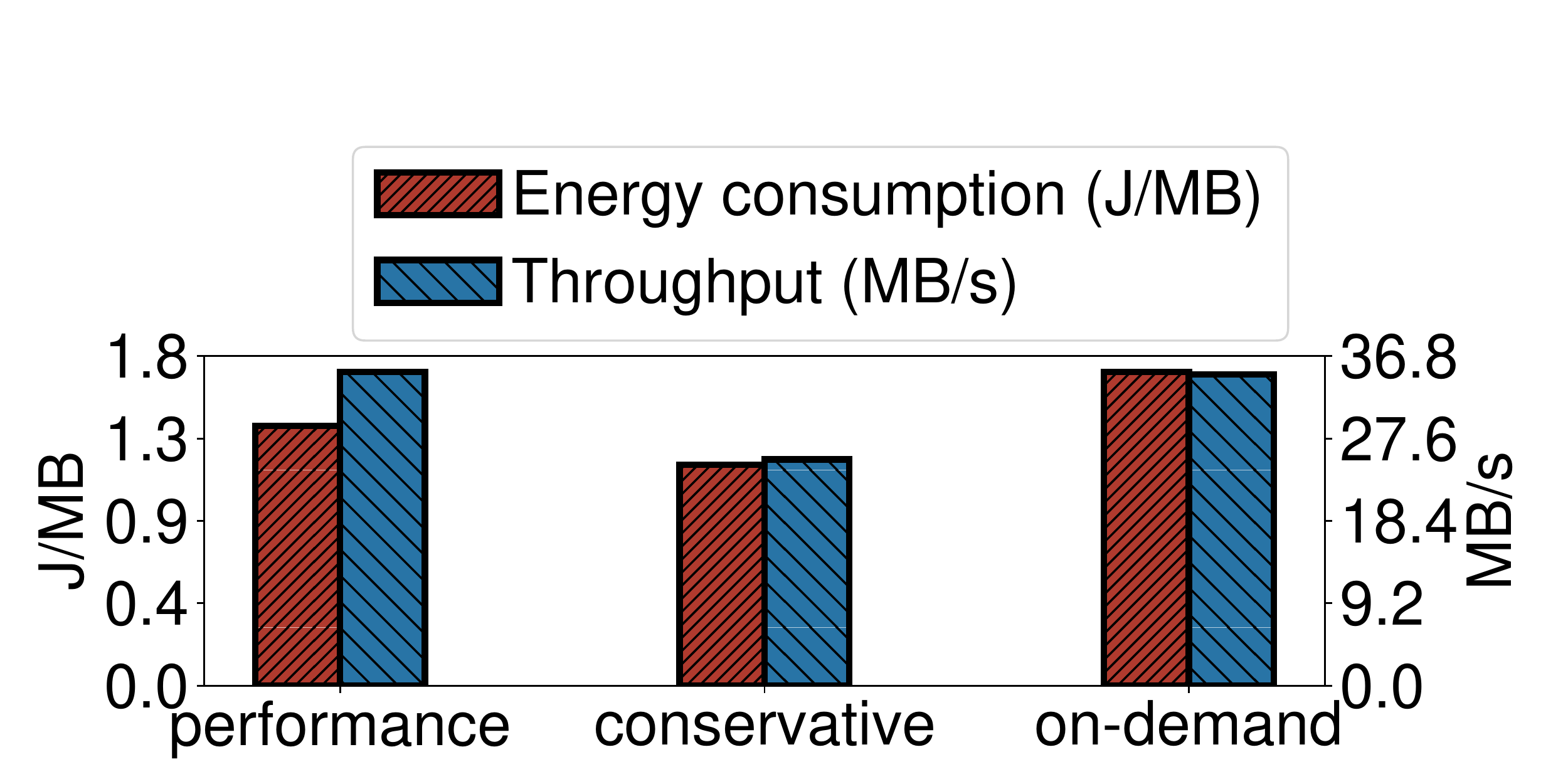}
        \caption{DVFS}
        \label{fig:freq_dvfs} 
    \end{subfigure}
    \caption{Impacts of frequency regulation for $Tcomp32$.}
\end{figure}

\subsection{Hardware Evaluation}
\label{subsec:evalu_hard}
In this section, we highlight how our novel framework, \design, explores various hardware design spaces, including architecture selection, ISA selection, frequency regulation, and core count adjustment. These are critical aspects in ensuring optimal energy consumption and throughput. We use the $Tcomp32$ algorithm and Rovio dataset as our primary test cases for this evaluation.

\subsubsection{Architecture Selection}
\design allows us to study the impact of different multicore configurations. We compare symmetric and asymmetric multicores under the same total computational power (i.e., maximum instructions per second). The roofline model benchmark \cite{roofline_1st,roofline_toolkit} shows that one 1.416GHz A72 big core in the RK3399 processor has about twice the computational power of one A53 little core at the same frequency (Figure\ref{fig:roofline}). Therefore, we compare different architecture configurations using the following settings:

\begin{itemize}
\item \textbf{Asymmetric multicore processor} (amp): Using 1 A72 big core and 2 A53 little cores at 1.416GHz.
\item \textbf{Symmetric multicore processor with only big cores} (smp\_big): Using 2 A72 big cores at 1.416GHz.
\item \textbf{Symmetric multicore processor with only little cores} (smp\_little): Using 4 A53 little cores at 1.416GHz.
\end{itemize}

Each architectural case is tuned to its maximum throughput, after which we compare their energy consumption along with throughput (Figure~\ref{fig:smp_amp}). The asymmetric multicore configuration (amp) outperforms the symmetric configurations (smp\_big and smp\_little), achieving both the lowest energy consumption and the highest throughput.

Furthermore, we observe that different stream compression steps (i.e., $s0\sim s2$ in Algorithm~\ref{alg:stateless}) involve varying \emph{operational intensities}\cite{amp_impacts_2005, amp_modeling, amp_scheduling_2021, amp_pie_2011} (i.e., instructions per memory access), as shown by the dashed lines in Figure\ref{fig:roofline}. This difference in operational intensities causes either over-provision or under-provision when conducting stream compression on symmetric multicores, thereby increasing energy consumption.

The $s0$ step leads to less performance gain when run on a big core than $s1$ and $s2$, as it primarily involves memory manipulation and makes out-of-order big cores over-provisioned. Therefore, much energy is wasted in the smp\_big configuration. Conversely, $s1$ and $s2$ are more worthwhile running on big cores as they offer enough computation density for big cores to support. Their high computation density also makes the little cores under-provisioned, resulting in energy dissipation in the smp\_little configuration. Through \design, we efficiently manage this architectural selection and intricacies, ensuring optimal energy usage and performance.

\subsubsection{ISA Selection}
\label{subsub:arch}

\design also enables us to investigate the effects of different Instruction Set Architectures (ISA) on stream compression performance. In this evaluation, we consider the RK3399, H2+, and Z8350 hardware platforms, as introduced in Section ~\ref{subsec:implentation_meter}.

Specifically, we compare the time and energy cost per core for each step of $Tcomp32$ under different ISAs (Figure~\ref{fig:unitTime} and ~\ref{fig:unitEnergy}). To account for their different working frequencies, we mathematically align their frequency to 1 GHz. The results for RK3399 are split into $RK3399_B$ and $RK3399_L$, representing its big and little cores, respectively.

Our evaluation highlights two key findings. First, the RISC ISA (used in RK3399) demonstrates superior performance compared to the CISC ISA (used in Z8350). Both RK3399\_B and RK3399\_L cores have significantly lower unit energy costs than the Z8350, and the RK3399\_B can even reduce the processing time of each step by 50\% compared to the Z8350. This is because RK3399's RISC-based execution hardware can directly execute the instructions used for stream compression, avoiding the extra overhead of micro-coding them. As a result, both unit latency and unit energy cost are reduced.

Second, traditional 32-bit processors (like the H2+) perform poorly in terms of both performance and energy efficiency. This is due to the limitations of a shorter register length. For example, manipulating a 33-bit intermediate result of $Tcomp32$ can be done with a single instruction on a single 64-bit register but requires two or more operations on 32-bit registers. This inefficiency at the instruction and register levels leads to significantly increased latency and is detrimental to energy efficiency.

From our analysis, we conclude that stream compression tasks should ideally be conducted on 64-bit RISC edge processors. These insights, derived from \design's evaluation capabilities, underscore the importance of choosing the right ISA for efficient stream compression on edge devices.

\subsubsection{Frequency Regulation}

\design provides flexibility in frequency regulation, allowing for both static frequency setting and dynamic voltage and frequency scaling (DVFS). In our evaluation, we explore these approaches' impact on throughput and energy consumption.

\textbf{Statically Tuning the clock frequency.} We first adjust the frequency of the big cores (denoted as ``B'') and the little cores (denoted as ``L'' ) on the RK3399 processor. We observe how frequency changes influence the throughput and energy consumption, as shown in Figure ~\ref{fig:freq_ene}. As expected, the relationship between frequency and throughput is nearly linear: higher frequency enables cores to execute more operations per unit time, and therefore more tuples are compressed.

Energy consumption, however, doesn't follow a simple monotonic relationship with frequency. It is the product of power and time, both of which respond differently to frequency changes. For example, a frequency of 0.408 GHz leads to lower power according to existing work~\cite{amp_DVFS1,amp_DVFS2}, but also involves more processing time, which counteracts the power reduction. Thus, it's less energy-efficient than the 0.6 GHz frequency. When the frequency surpasses 0.816GHz, the energy consumption increases with frequency because the rise in power is more significant than the time reduction.

\textbf{DVFS.} We also employ the DVFS approach~\cite{amp_DVFS1,amp_DVFS2,amp_DVFS3} to dynamically adjust the frequency. We use different DVFS strategies and present the results in Figure~\ref{fig:freq_dvfs}. The ``performance'' strategy, which fixes each core at its highest frequency without dynamic reconfiguration (1.8GHz for big cores and 1.416GHz for little cores), serves as a reference. The ``conservative'' and ``on-demand'' strategies attempt to reduce energy consumption by dynamically reconfiguring frequency. The primary difference is that the ``conservative'' strategy changes frequency less frequently than ``on-demand''.

Our results indicate that the ``conservative'' strategy can further reduce energy consumption by 15\% compared to the default ``performance strategy'', albeit at the cost of a 38\% increase in latency. This strategy offers a coarser-grained balance of energy efficiency and latency constraints, since the overhead of dynamic frequency regulation can introduce latency variability. In contrast, the ``on-demand'' strategy doesn't provide any benefits; it actually increases both latency and energy consumption due to the high overhead of frequent frequency switches.

Through this exploration, \design highlights the critical role of frequency regulation in achieving energy-efficient stream compression, guiding us towards more efficient hardware configurations and settings.

\subsubsection{Tuning the Number of Cores}
Finally, we demonstrate the ability of \design to effectively manage the number of cores for stream compression tasks. 
\setlength{\columnsep}{-1pt}%
\begin{wrapfigure}[9]{r}{0.48\columnwidth}
    \centering
    \raisebox{-2pt}[\dimexpr\height-1.2\baselineskip\relax]{\includegraphics*[width=0.5\columnwidth]{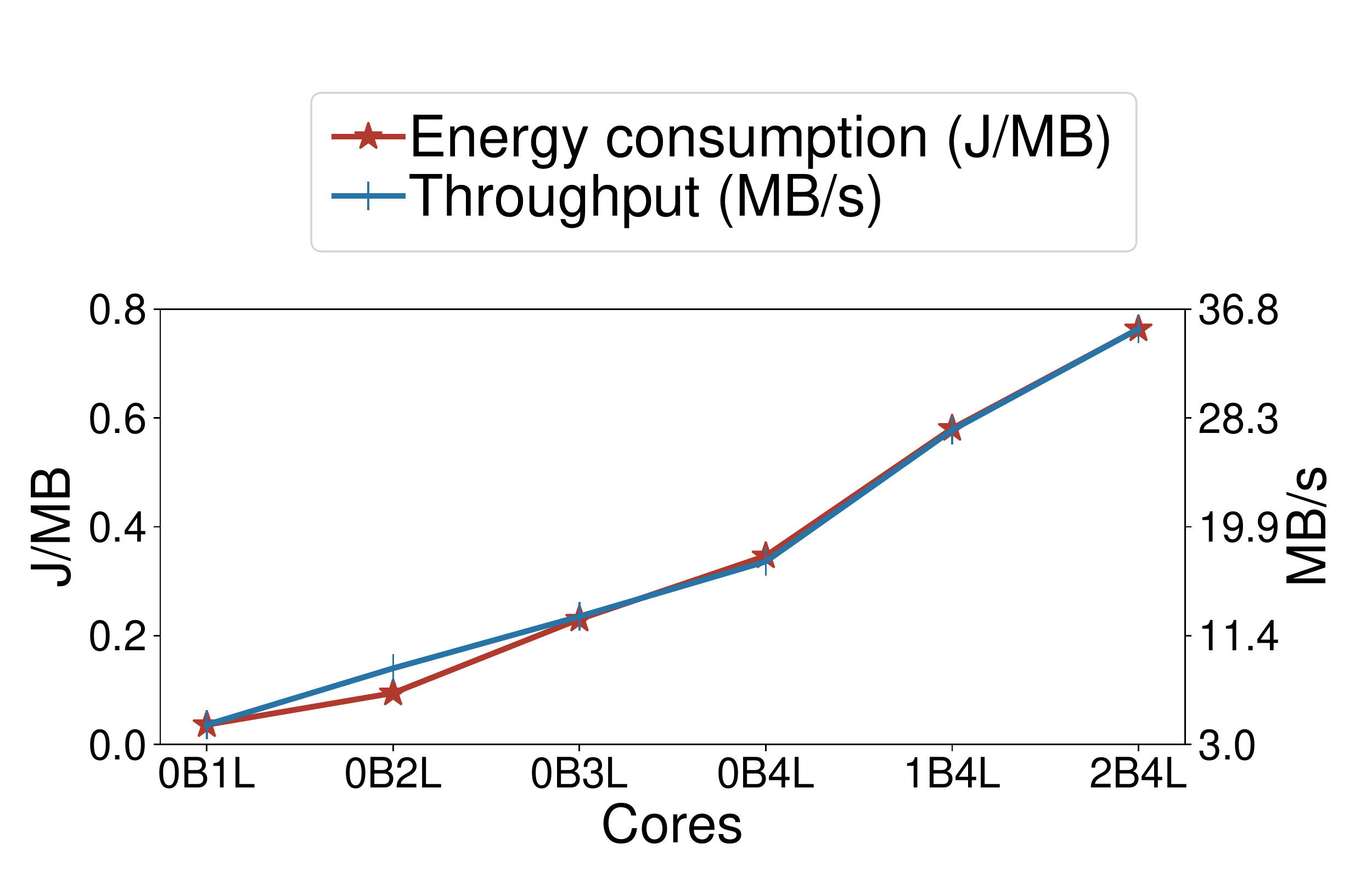}}
    \caption{Scalability} \label{fig:cores}
\end{wrapfigure}
The results are shown in Figure~\ref{fig:cores}. By enabling different numbers of big and little cores, we observe a trade-off between energy consumption and throughput. This demonstrates \design's flexible core management strategy, striking a balance between energy efficiency and throughput.


\subsection{Parallelization Strategies}
\label{subsec:evalu_para}
We now explore various parallelization strategies provided by \design, including different execution strategies, micro-batch size, state sharing, and scheduling strategies. Our exploration, illustrated with the $Tcomp32$ algorithm and the Rovio dataset, illustrates the importance of each strategy and its impacts on energy consumption and throughput.
 
\subsubsection{Execution Strategy: Eager vs Lazy}
\label{subsub:batch}
A key component of our \design framework is the ability to vary the execution strategy. Specifically, we explore the differences between eager and lazy execution strategies using the $Tcomp32$ algorithm and the Rovio dataset.

\begin{figure}[t!]
    \centering
    \begin{subfigure}[c]{0.46\columnwidth}
        \centering
        \includegraphics[width=\columnwidth]{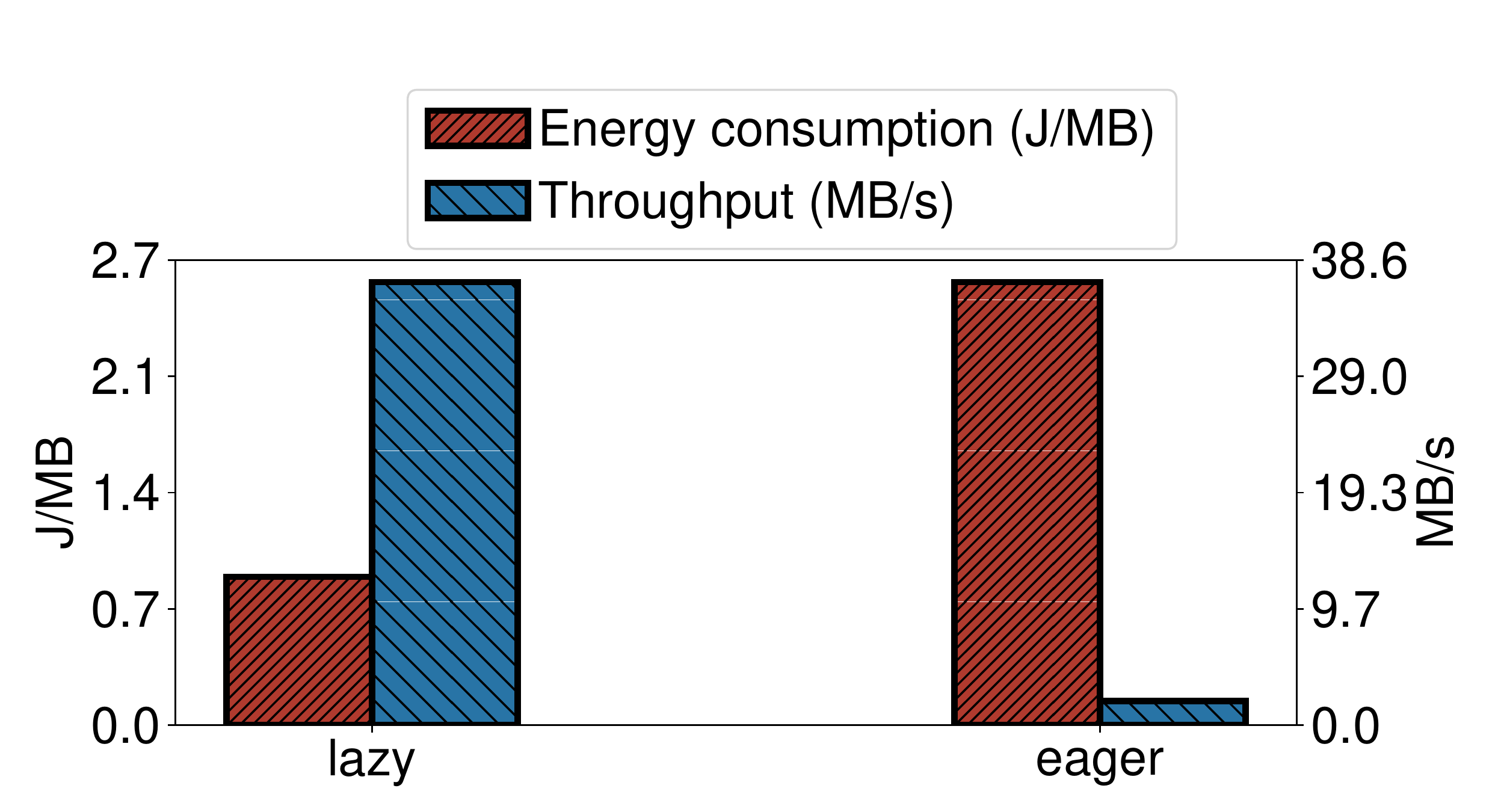}
        \caption{Energy and throughput consumption}
        \label{fig:lazy_eager}
    \end{subfigure}%
    \begin{subfigure}[c]{0.63\columnwidth}
        \centering        
        \resizebox{.7\columnwidth}{!}{%
        \begin{tabular}{|p{1.4cm}|p{1.3cm}|p{1.3cm}|}
        \hline
        Execution way& Blocked time& Running time                           \\ \hline
         Eager  & 0.53         & 0.03                                   \\ \hline
         Lazy   &  0.006        &  0.03                               \\ \hline
        \end{tabular}
        }
        \caption{Processing time breakdown}
        \label{tab:blk_run}
    \end{subfigure}
    \caption{Impacts of execution strategy for $Tcomp32$.}
\end{figure}

Under eager execution, each incoming tuple is compressed as soon as it arrives. On the contrary, under lazy execution (implemented as the default setting in \design), compression is conducted only after the accumulation of a 400-byte micro-batch. These strategies do not affect the compression ratio but significantly influence throughput and energy consumption.

Figure~\ref{fig:lazy_eager} clearly illustrates the superiority of the lazy strategy, leading to higher throughput and lower energy consumption. The eager approach incurs a high energy cost because it must immediately distribute each incoming tuple to the running cores, thus reducing parallelism. In contrast, the lazy strategy enhances parallelism by accumulating workloads before distribution.

To better understand the cost difference between the two execution strategies, we break down the processing time into 'blocked' and 'running' phases, as presented in Table~\ref{tab:blk_run}. 'Blocked' time is spent resolving concurrent read/write conflicts and mitigating cache trashing, while 'running' time is devoted to the actual compression process. The eager execution strategy incurs significantly more blocked time, leading to high overhead and reduced energy efficiency.

\begin{figure}[t!]
    \centering
    \begin{subfigure}[t]{0.5\columnwidth}
        \centering
        \includegraphics[width=\columnwidth]{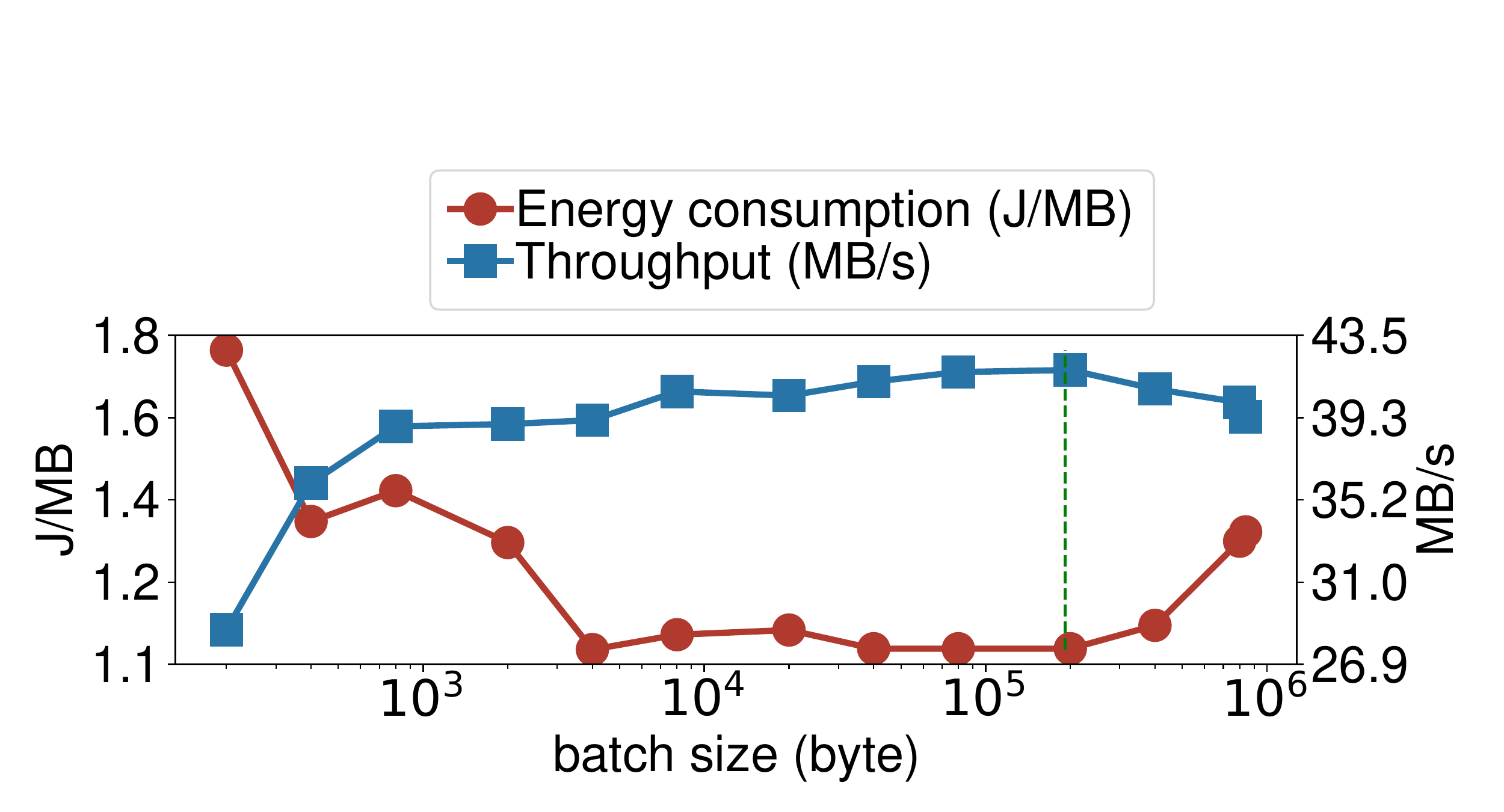}
        \caption{Energy and throughput consumption}
        \label{fig:impact_batch}
    \end{subfigure}%
    ~ 
    \begin{subfigure}[t]{0.5\columnwidth}
        \centering
        \includegraphics[width=\columnwidth]{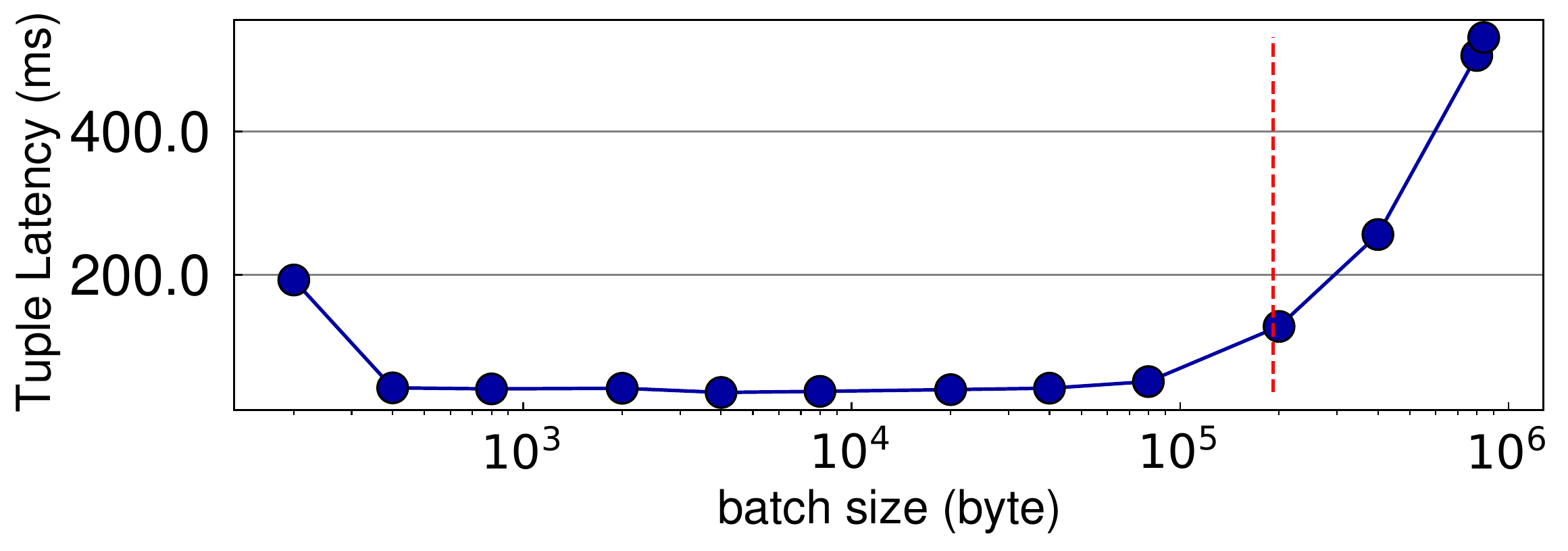}
        \caption{Average latency; The red dashed line is the total L1D size of all cores}
        \label{fig:impact_batch_lat}
    \end{subfigure}
    \caption{Impacts of batch sizes for $Tcomp32$.}
\end{figure}

\textbf{Batch Size.} 
In our study of execution strategies, we further investigate the effects of varying batch sizes during lazy execution. Continuing with the $Tcomp32$ algorithm and Rovio dataset, we vary the batch size from hundreds of bytes to millions of bytes. The impacts of batch size on energy consumption and throughput are illustrated in Figure~\ref{fig:impact_batch}.

The results show that both minimum energy consumption and maximum throughput occur simultaneously at an optimal, moderate batch size. Any deviation from this size—either by increasing or decreasing the batch—leads to reduced throughput and increased energy consumption. This finding underscores the importance of optimizing batch size for efficient stream compression on edge devices.

Significantly, we draw attention to a correlation between the optimal batch size and the total L1D cache size of all cores, indicated by the red dashed line in Figure~\ref{fig:impact_batch}. The closeness of these values suggests that efficient micro-batching should be cognizant of the L1D cache size. Hence, \design's strategy for lazy execution, which includes an L1D-cache-aware approach to micro-batching, can offer substantial advantages in terms of energy efficiency and throughput.

As we continue to evaluate the implications of the batch size in our lazy execution strategy, we now turn our attention to the role of latency. Using the $Tcomp32$ algorithm and the Rovio dataset, we consider average latency under different batch sizes. These results are presented in Figure~\ref{fig:impact_batch_lat}.

Interestingly, we observe a curve where latency first decreases and then increases as the batch size changes. This inverted U-shape pattern implies a trade-off in determining the optimal batch size. It is crucial to balance the competing requirements of lower latency and improved throughput and energy efficiency.

Furthermore, the increase in latency appears to correlate strongly with the total L1D cache size of all cores, highlighted by the red dashed line in Figure~\ref{fig:impact_batch_lat}. Upon exceeding this cache size, latency sees a noticeable increase, which can be attributed to the higher likelihood of cache misses as the batch size outstrips the available cache.


These observations reinforce the importance of an L1D-cache-aware approach to setting the batch size in \design's lazy execution strategy. Balancing the demands of latency, throughput, and energy efficiency requires a sophisticated approach to micro-batching — a challenge that \design effectively navigates with its adaptive and hardware-conscious design.

\begin{figure}[t!]
    \centering
    \begin{subfigure}[t]{0.5\columnwidth}
        \centering
        \includegraphics[width=\columnwidth]{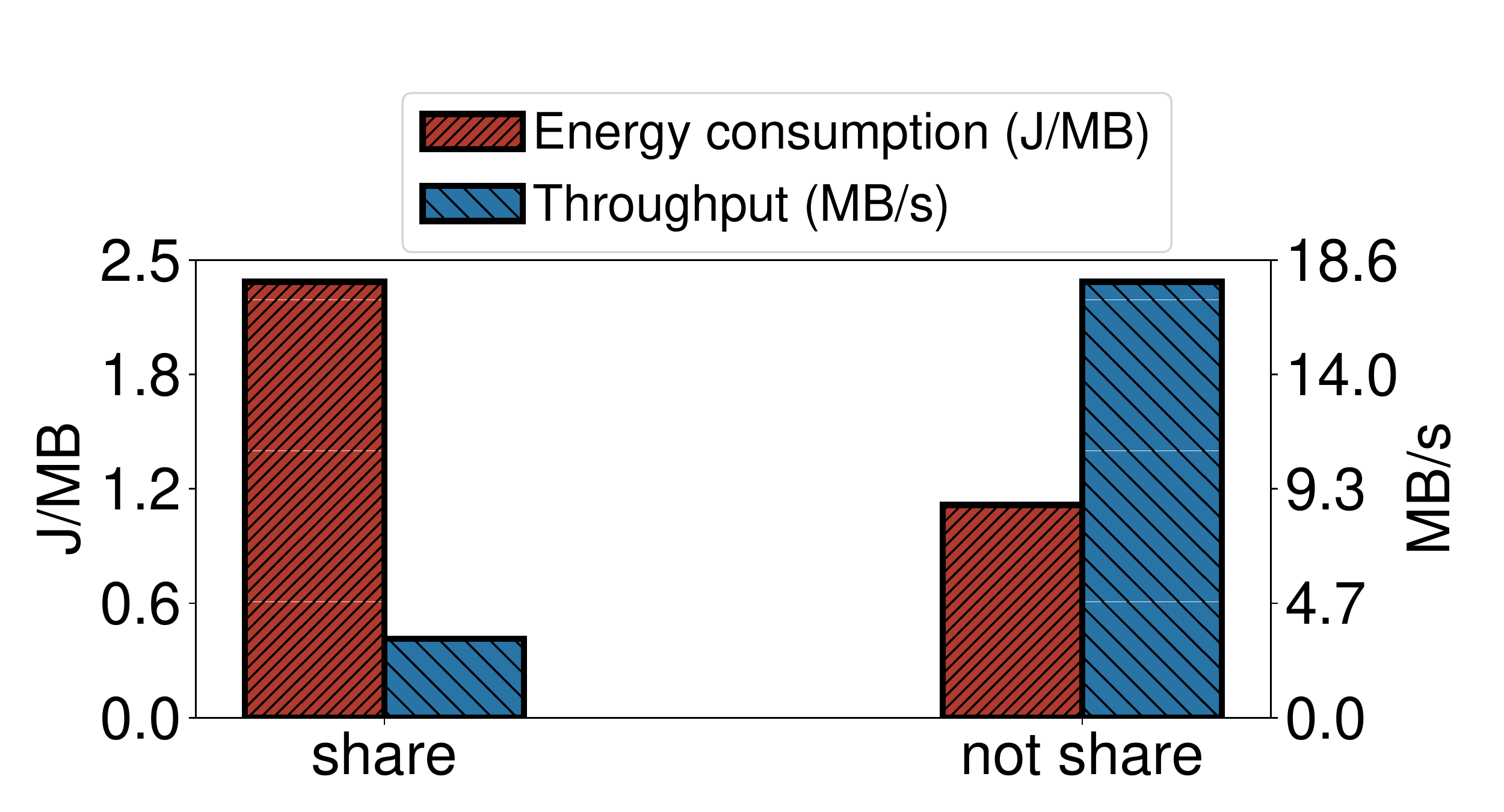}
        \caption{Energy and throughput comparison}
        \label{fig:cost_stateShare}
    \end{subfigure}%
    ~ 
    \begin{subfigure}[t]{0.5\columnwidth}
        \centering
        \includegraphics[width=\columnwidth]{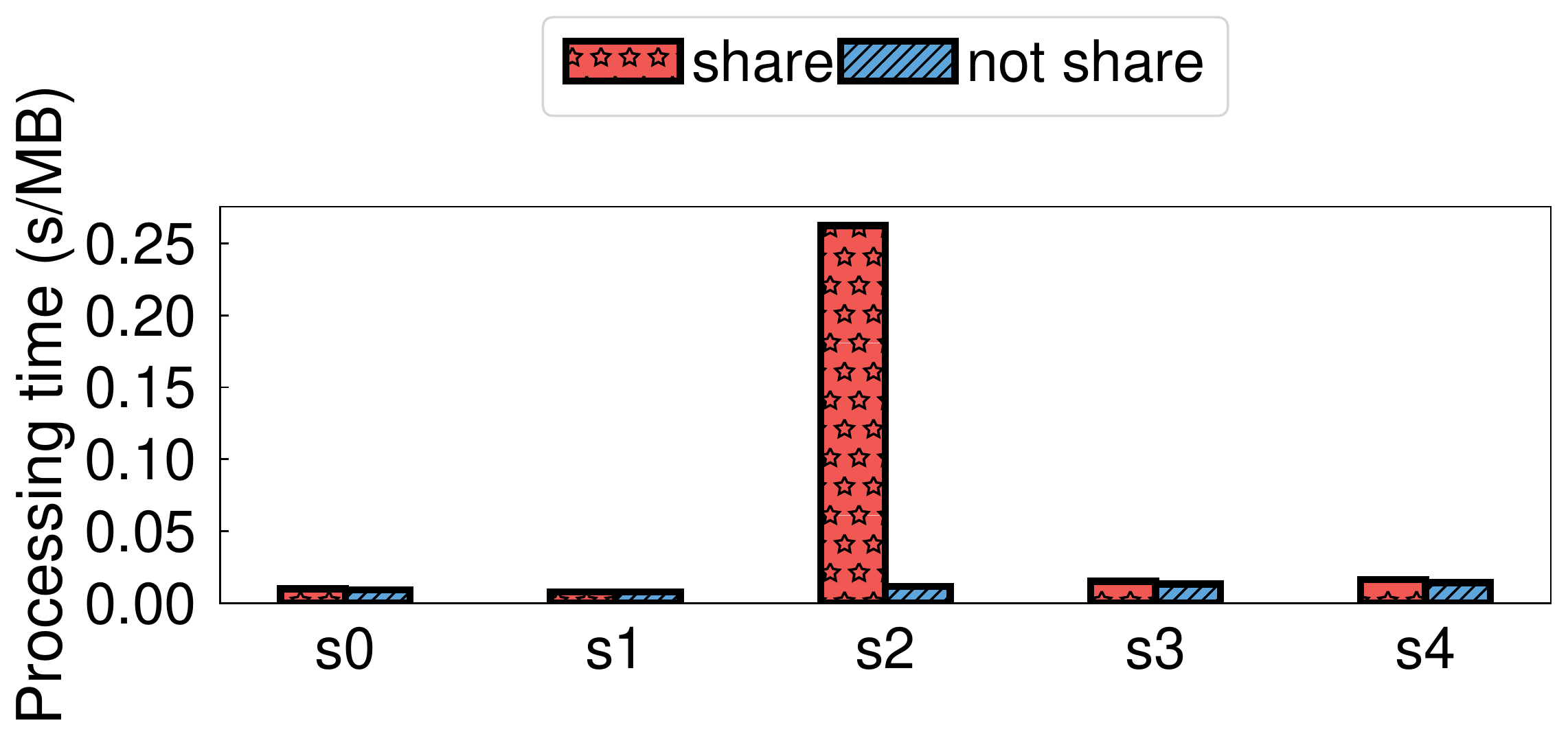}
        \caption{Step breakdown comparison}
        \label{fig:breakdown_stateful} 
    \end{subfigure}
    \caption{Impacts of state management strategy for $Tdic32$.}
\end{figure}
\subsubsection{State Management Strategy: Shared vs Private}
\label{subsub:statemgmt}
One of the critical considerations within the realm of stateful stream compression in \design is the choice between sharing the state across threads or maintaining private states for each thread. We utilized the $Tdic32$ algorithm and the Rovio dataset to investigate the trade-offs associated with each state management strategy.

In the shared state scenario, each working thread shares a common concurrent dictionary. In contrast, with private state management, each thread maintains its private dictionary. These two strategies are elaborated further in Section~\ref{subsec:intro_state}. Figure~\ref{fig:cost_stateShare} compares the resulting energy consumption and throughput for each scenario.

Interestingly, state sharing incurs a significantly higher energy consumption and concurrently reduces throughput. However, the shared state strategy only marginally improves the compression ratio, moving from $1.78$ without state sharing to $1.81$ with state sharing. In most cases, pursuing a mere 3\% improvement in compression ratio is not worthwhile, especially considering the overhead involved.

\textbf{Execution Time Breakdown.} To delve deeper into the source of the increased overhead with shared state management, we broke down the processing time for each step, as defined in Algorithm~\ref{alg:stateful}, for the two implementation versions. This breakdown is presented in Figure~\ref{fig:breakdown_stateful}.

The analysis revealed that the bulk of the added cost is accrued during the state updating step (i.e., $s2$), where frequent locking significantly reduces parallelism. The cores, while waiting for locks, are mainly engaged in memory access, which is less energy-demanding than arithmetic calculations. This explains why the rise in energy consumption is not as pronounced as the drop in throughput with shared state management. 

This in-depth analysis underscores the preference for a private state management strategy in \design's stateful stream compression, enhancing throughput and energy efficiency without significantly compromising the compression ratio.

\subsubsection{Varying Scheduling Strategy}
\begin{figure}[t!]
    \centering
    \begin{subfigure}[t]{0.5\columnwidth}
        \centering
        \includegraphics[width=\columnwidth]{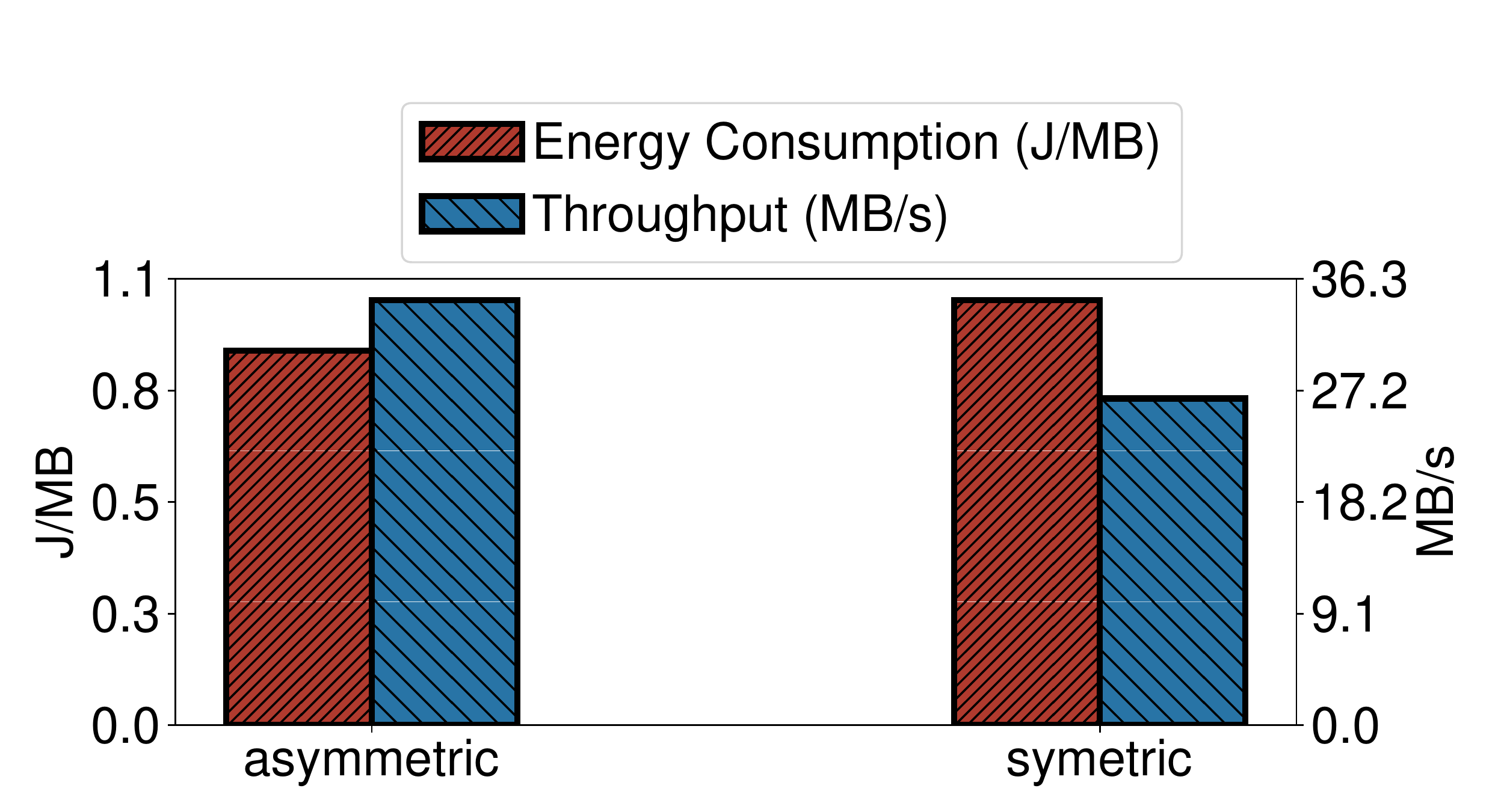}
        \caption{Varying scheduling strategy}
        \label{fig:scheduling}
    \end{subfigure}%
    ~ 
    \begin{subfigure}[t]{0.5\columnwidth}
        \centering
        \includegraphics[width=\columnwidth]{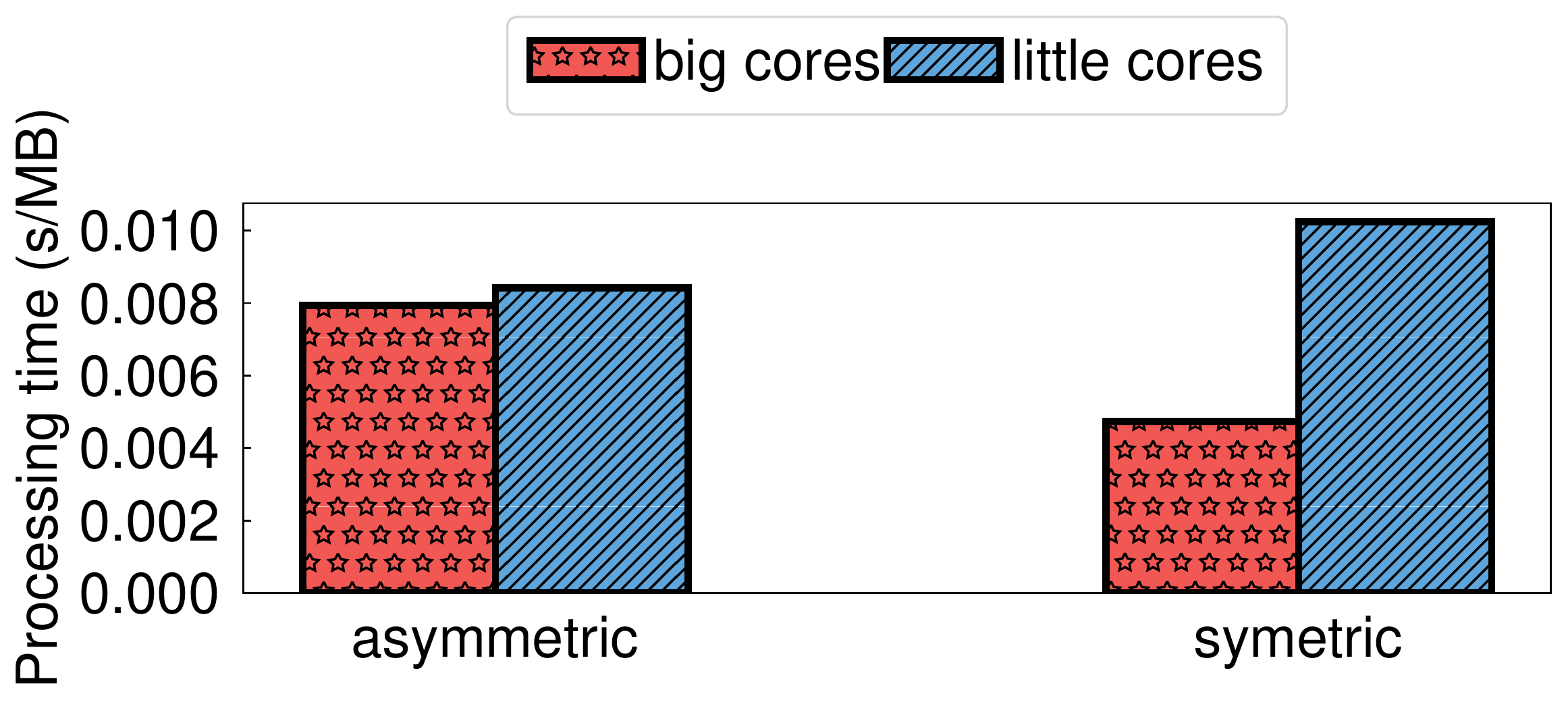}
        \caption{Time breakdown}
        \label{fig:scheduling_bd} 
    \end{subfigure}
    \caption{Impacts of scheduling strategy for $Tcomp32$.}
\end{figure}

In order to determine the optimal scheduling strategy, we explore both symmetric and asymmetric approaches using the $Tcomp32$ algorithm and the Rovio dataset. As the scheduling strategy doesn't influence the compression ratio, we focus our evaluation on energy consumption and throughput, as presented in Figure~\ref{fig:scheduling}.
Interestingly, symmetric scheduling results in both a reduction in throughput (26.2\%) and an increase in energy consumption (13.4\%) compared with asymmetric scheduling. The underlying reason is that symmetric scheduling fails to take into account the differences in hardware, as we have previously illustrated in Figure~\ref{fig:roofline}. This oversight leads to a wastage of the superior computational power of big cores.

For instance, Figure~\ref{fig:scheduling_bd} shows the processing time for the $s1$ step of the $Tcomp32$ algorithm for both big and little cores. It becomes evident that under symmetric scheduling, big cores spend a considerable amount of time waiting for little cores. This inefficient waiting period is the primary contributor to the sub-optimal energy efficiency and performance under symmetric scheduling.

To circumvent this issue, we strongly recommend asymmetric scheduling. This approach respects the unique computational capabilities of different cores, efficiently allocating tasks in a manner that optimizes performance and energy usage. As a result, it achieves higher throughput and lower energy consumption than symmetric scheduling, demonstrating its superiority in managing heterogeneous multi-core systems for stream compression tasks.
\begin{figure}[t!]
    \centering
    \begin{subfigure}[t]{0.5\columnwidth}
        \centering
        \includegraphics[width=\columnwidth]{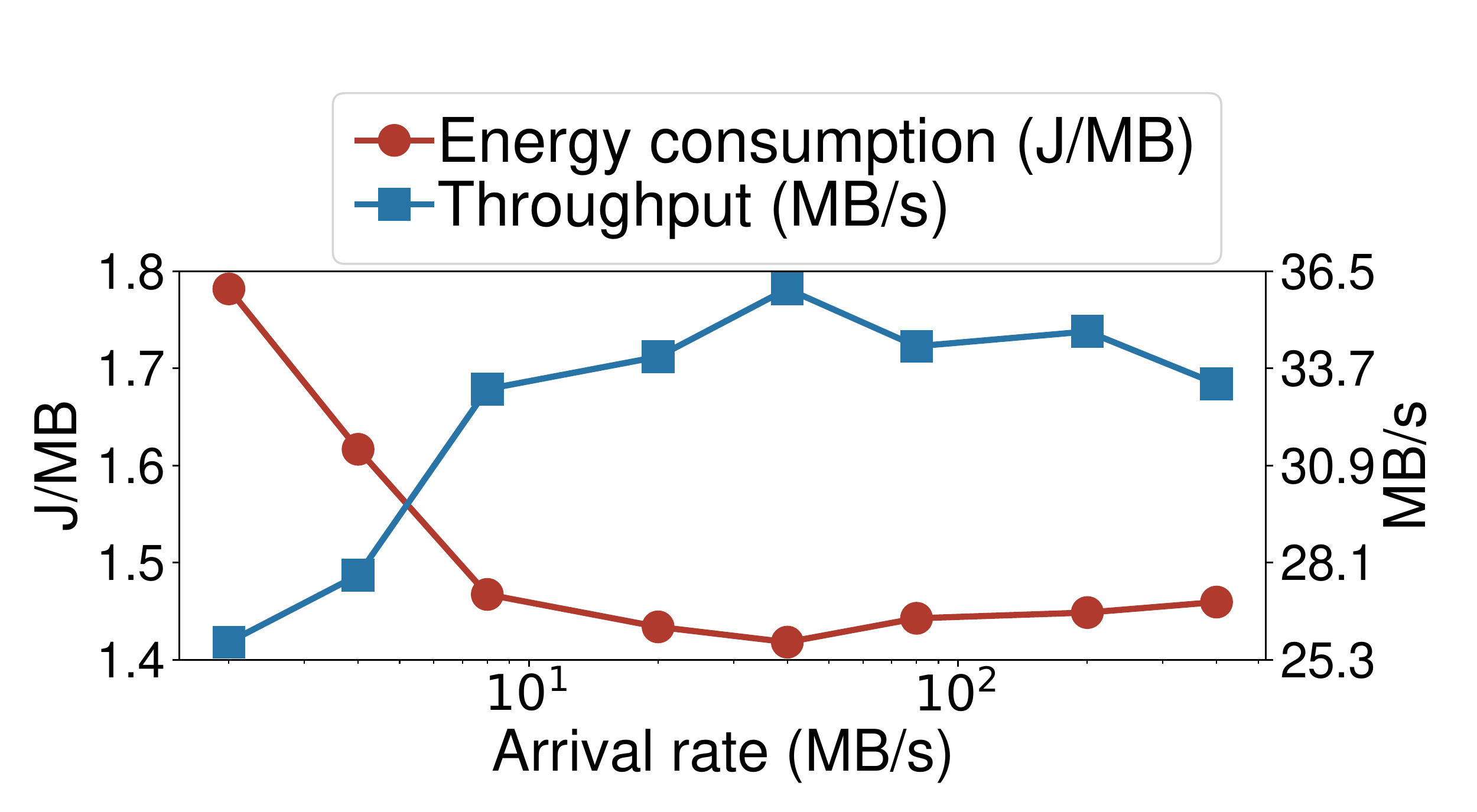}
        \caption{Arrival rate}
         \label{fig:impact_rate}   
    \end{subfigure}%
    ~ 
    \begin{subfigure}[t]{0.5\columnwidth}
        \centering
        \includegraphics[width=\columnwidth]{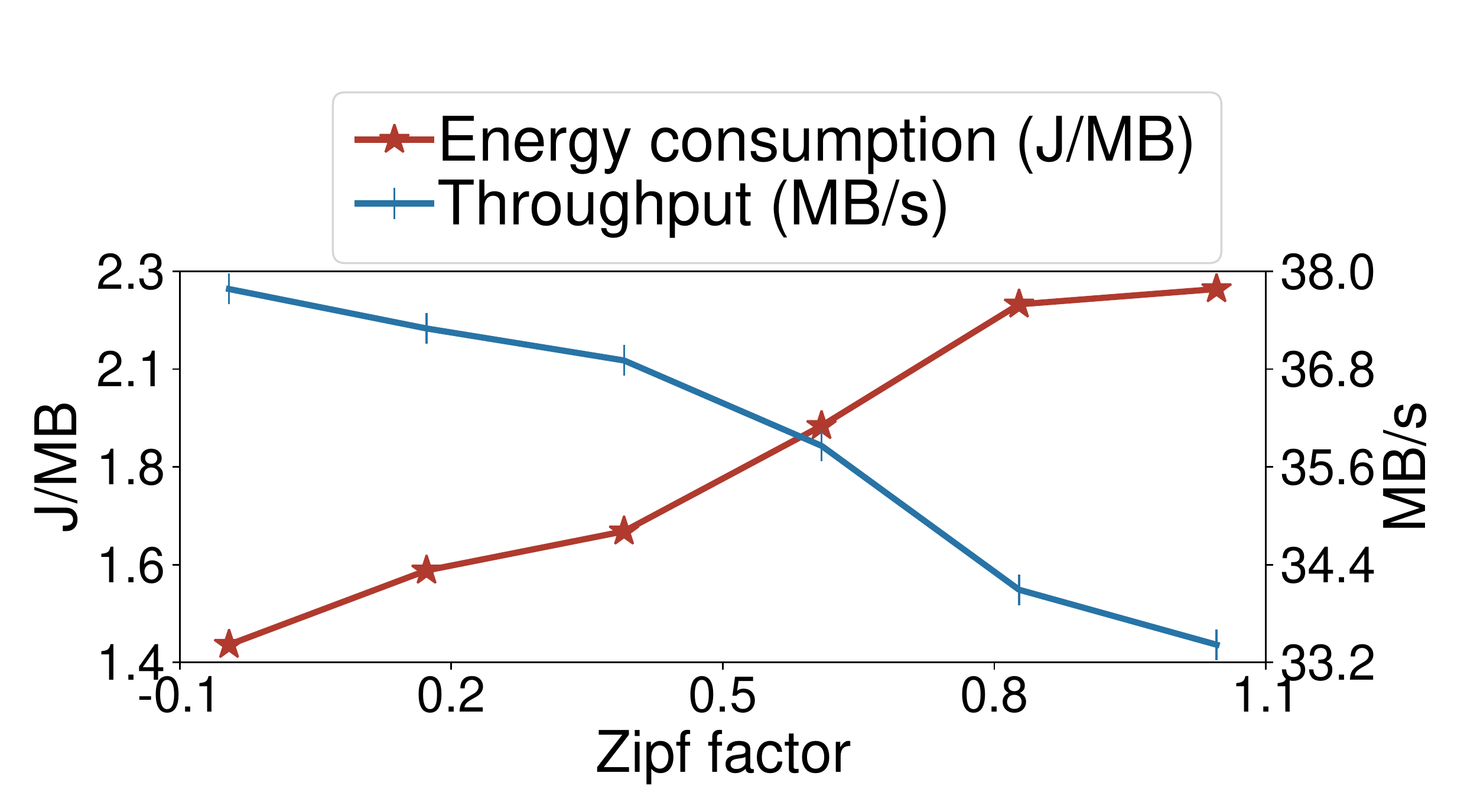}
        \caption{Arrival skewness}
         \label{fig:impact_skewness}     
    \end{subfigure}
    \caption{Impacts of arrival pattern for $Tcomp32$.}
\end{figure}

\subsection{Workload Sensitivity Study}
\label{subsec:evalu_workload}
In this subsection, we illustrate the robustness of our framework, \design, by conducting a comprehensive sensitivity analysis based on varying IoT data workloads. Our goal is to assess how \design responds under diverse arrival patterns and differing levels of data compressibility. Our evaluation utilizes the $Tcomp32$ algorithm to compress the Rovio dataset while adjusting the arrival pattern of tuples, and a synthetic dataset, Micro, for evaluating the impact of data compressibility.

\subsubsection{Arrival Pattern}
Arrival patterns of tuples can significantly affect the performance of stream compression algorithms. By manipulating the order of timestamps in the Rovio dataset, we are able to modify the tuple arrival characteristics while keeping other parameters constant.

\textbf{Impacts of arrival rate.} The impact of varying arrival rates is depicted in Figure~\ref{fig:impact_rate}. Here, the arrival rate of tuples varies from $500$ to $10^6$ per second. When the arrival rate is low, the hardware is underutilized, resulting in increased processing latency. This underutilization is due to more cycles of periodical DDR4 refreshing~\cite{amp_ddr4}, leading to increased overhead. Conversely, when the arrival rate is high, processing latency increases due to resource conflicts, such as cache misses.

\textbf{Impacts of arrival skewness.} Figure~\ref{fig:impact_skewness} shows the impact of varying levels of arrival skewness on the latency of $Tcomp32$. Arrival skewness is varied by adjusting the Zipf factor from 0 to 1. Increased skewness leads to higher latencies as both under-utilization and overloading cases become more prevalent.

\subsubsection{Data Compressibility}
We further assess the impact of data compressibility on the performance of both stateless and stateful compression algorithms using the Micro dataset. The stateless $Tcomp32$ and stateful $Tdic32$ compression algorithms are chosen for this analysis.
\begin{figure}[t!]
    \centering
    \begin{subfigure}[t]{0.5\columnwidth}
        \centering
        \includegraphics[width=\columnwidth]{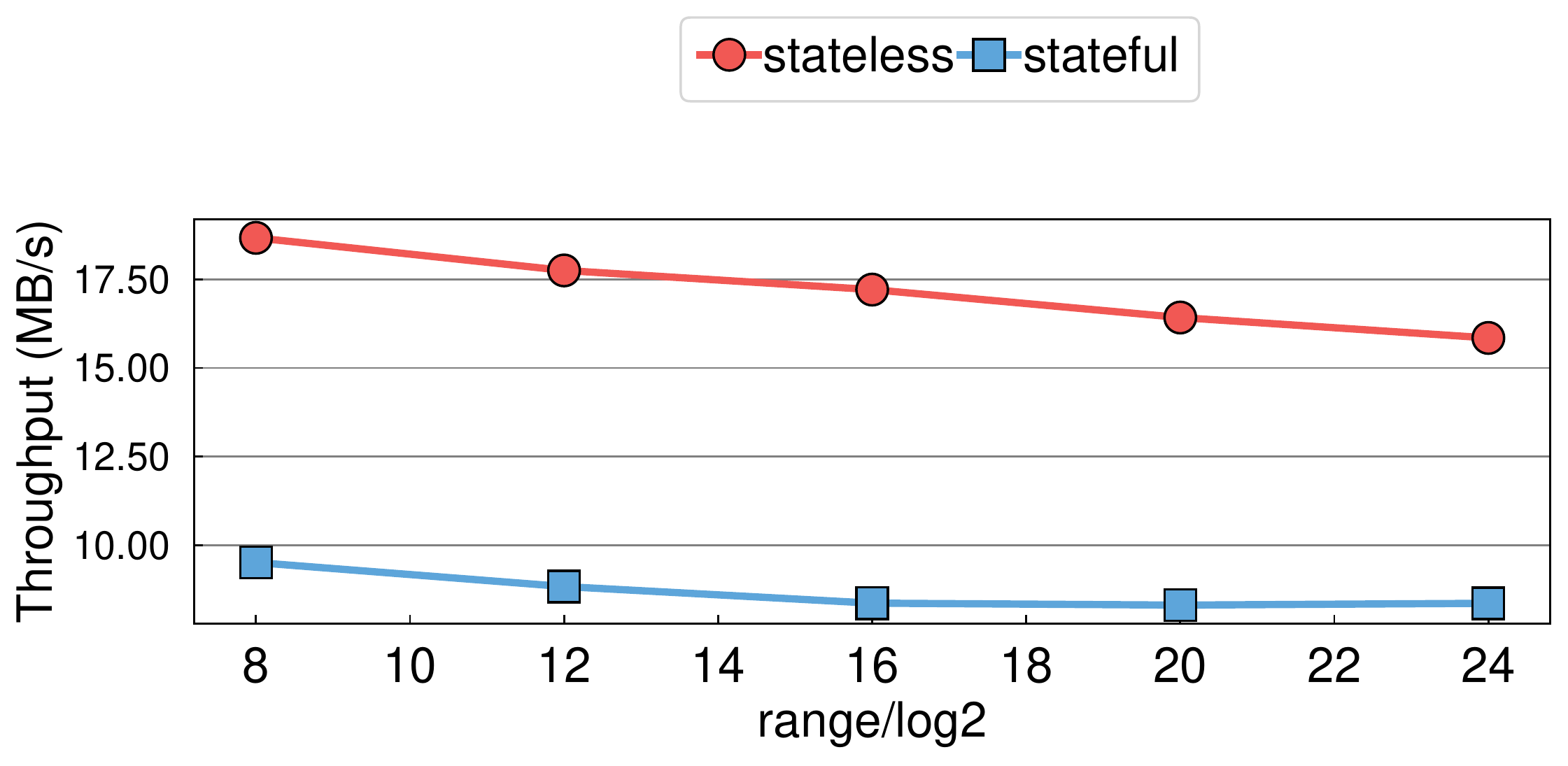}
        \caption{Throughput}
         \label{fig:impact_range_thr} 
    \end{subfigure}%
    ~ 
    \begin{subfigure}[t]{0.5\columnwidth}
        \centering
        \includegraphics[width=\columnwidth]{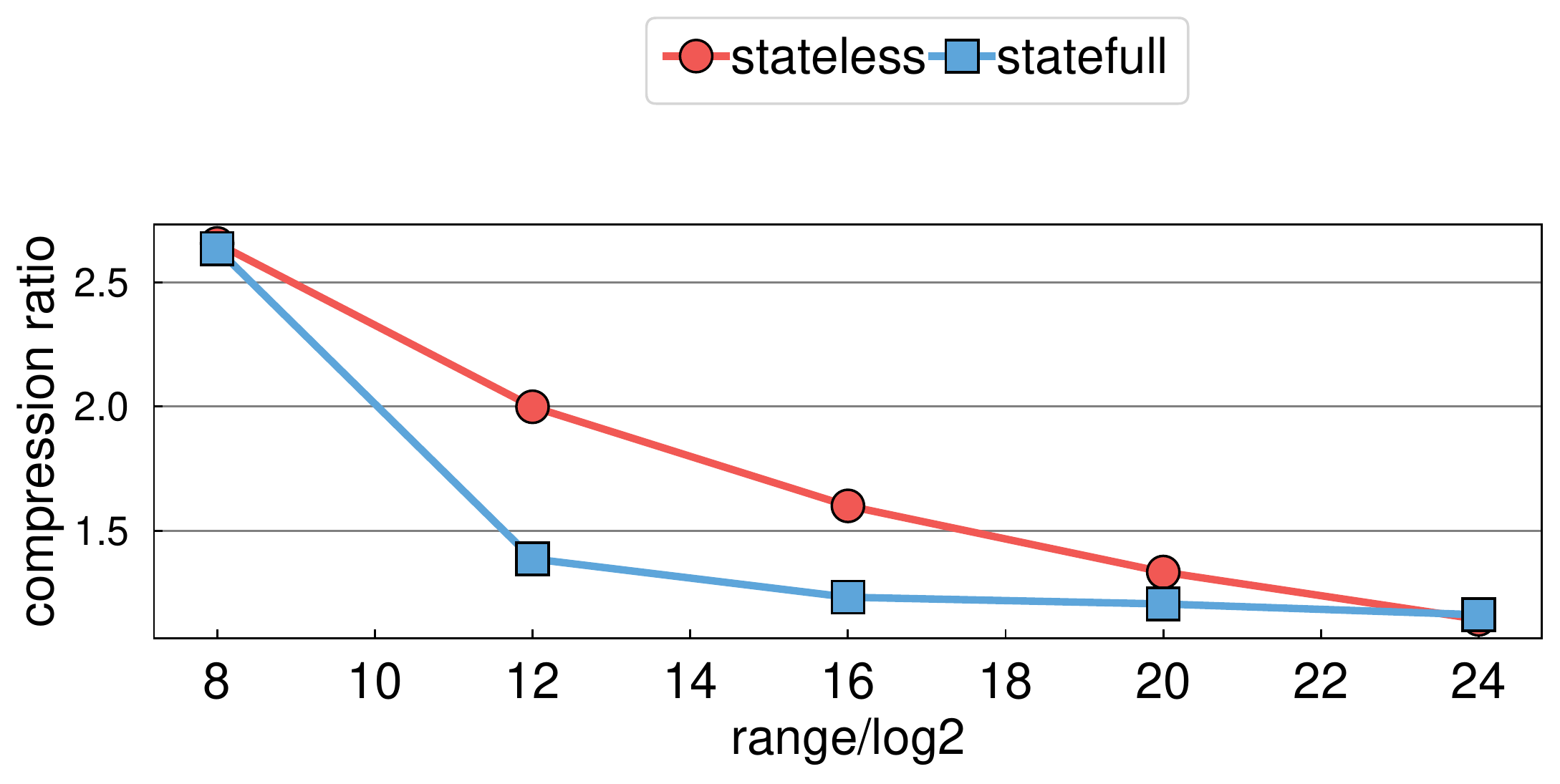}
        \caption{Compression ratio}
        \label{fig:impact_range_ratio}  
    \end{subfigure}
    \caption{Impacts of dynamic range for stateless ($Tcomp32$) and stateful stream compression ($Tdic32$).}
\end{figure}

\textbf{Impacts of stateless compressibility.} Stateless compressibility can be exploited by analyzing a single piece of input data without referring to a state. Figure~\ref{fig:impact_range_thr} and ~\ref{fig:impact_range_ratio} show the throughput and compression ratio respectively for varying levels of stateless compressibility. For the stateless $Tcomp32$, as the dynamic range increases, it becomes more costly and less compressible. However, $Tdic32$ exhibits a "cliff effect" at around $2^{12}$, corresponding to its dictionary entries. Before this threshold, $Tdic32$ achieves higher compressibility and lower latency, after which the compressibility becomes nearly constant.

\begin{figure}[t!]
    \centering
    \begin{subfigure}[t]{0.5\columnwidth}
        \centering
        \includegraphics[width=\columnwidth]{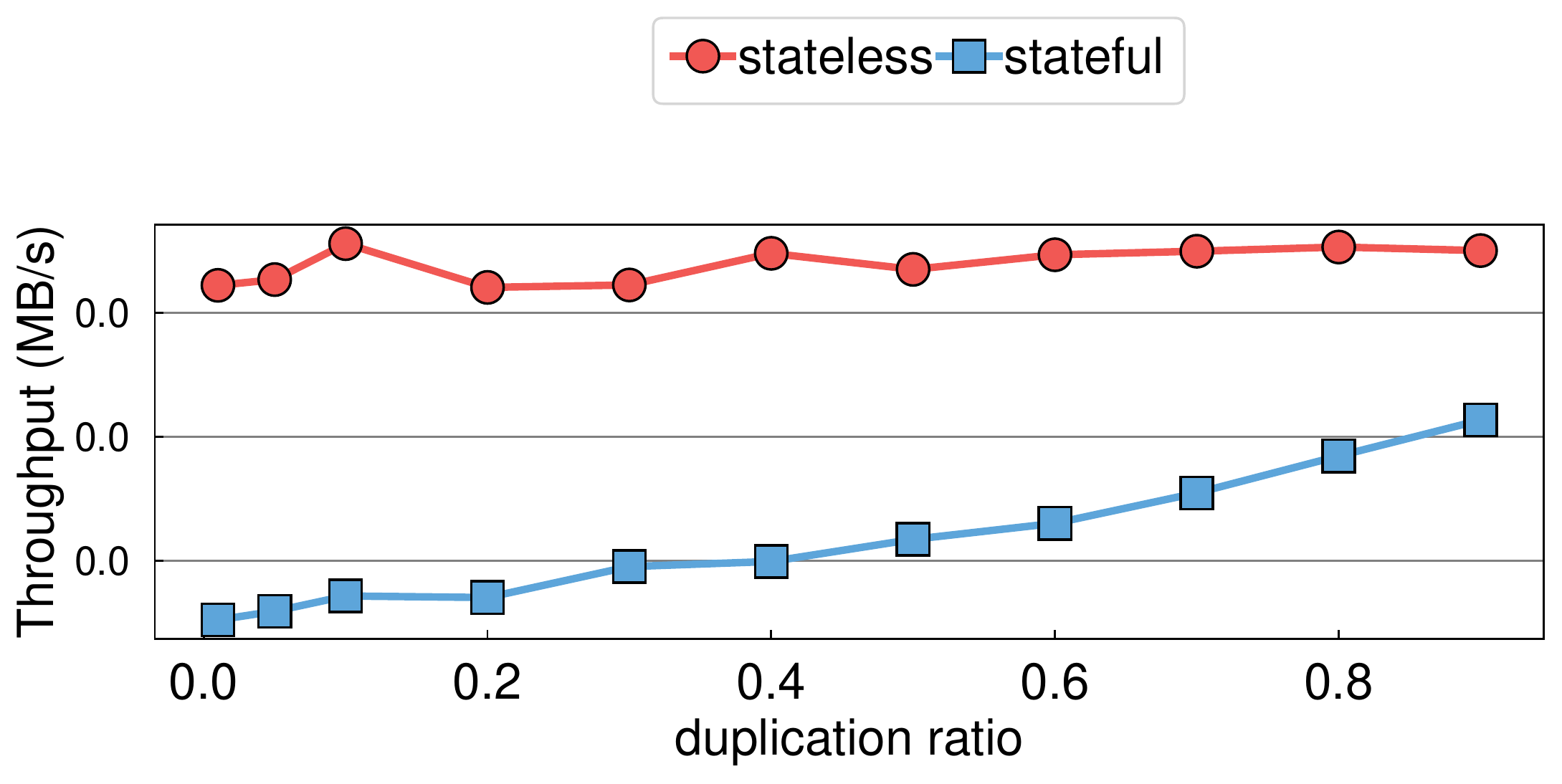}
        \caption{Throughput}
         \label{fig:impact_duplication_thr} 
    \end{subfigure}%
    ~ 
    \begin{subfigure}[t]{0.5\columnwidth}
        \centering
        \includegraphics[width=\columnwidth]{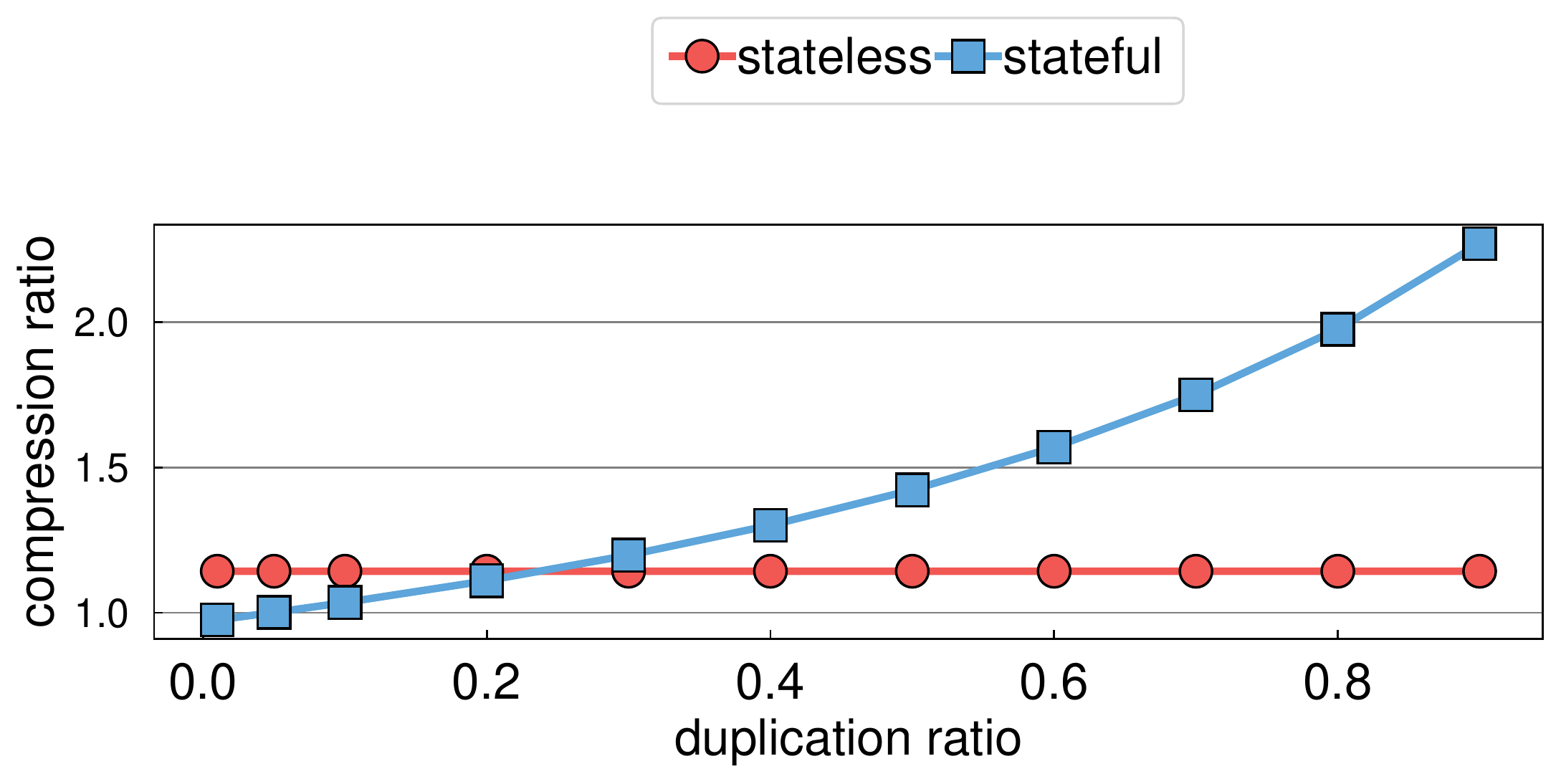}
        \caption{Compression ratio}
        \label{fig:impact_duplication_ratio}
    \end{subfigure}
    \caption{Impacts of tuple duplication for stateless ($Tcomp32$) and stateful stream compression ($Tdic32$).}
\end{figure}

\textbf{Impacts of stateful compressibility.} Stateful compressibility, on the other hand, can be detected by referencing the history of the compression process. Figure~\ref{fig:impact_duplication_thr} and~\ref{fig:impact_duplication_ratio} show the throughput and compression ratio respectively for varying levels of stateful compressibility. As the duplication ratio increases, the stateful $Tdic32$ experiences higher throughput and a higher compression ratio due to less frequent state updates and fewer bits of compressed data output. In comparison, the stateless $Tcomp32$ is largely unaffected by changes in stateful compressibility.

\subsection{Summary}
\label{subsec:summary}
Our comprehensive investigation and evaluations have elucidated a number of key insights that underline the efficacy of our novel \design framework, and are summarized below:

Firstly, stream compression algorithm diversity: We found that no single stream compression algorithm is universally superior. The choice of algorithm greatly depends on the specific application requirements and the nature of the workload. Even though the trade-offs involved in algorithm selection are complex, we discovered that lossy stream compression consistently achieves the highest compression ratio across all real-world scenarios with a negligible compromise on information integrity. Notably, \design's versatility allows for the seamless integration of various compression algorithms based on the unique requirements of each IoT application.

Secondly, parallelization design: Efficient parallelization is paramount in stream compression, and \design ensures this through its strategic design principles. It judiciously determines the parallelization granularity and employs a cache-aware agglomeration strategy. The absence of such considerations could lead to a significant penalty, up to 11x in some cases.

Thirdly, embracing asymmetric 64-bit RISC edge processors: Our research also underscores the significant advantages of performing stream compression on asymmetric 64-bit RISC edge processors, which can lead to substantial improvements in latency and energy efficiency. \design is specifically designed to be compatible with these processors and utilizes their frequency and core regulation mechanisms to maximize efficiency.

In essence, the success of an IoT stream compression system hinges on its ability to 1) dynamically adapt to the IoT workload and optimally choose the compression approach, 2) strategically determine parallelization granularity and utilize cache-aware agglomeration, and 3) be compatible with asymmetric 64-bit RISC edge processors and regulate their resources efficiently based on varying user demands. With its versatility, adaptability, and resource efficiency, \design satisfies these requirements and sets a new benchmark in stream compression for IoT analytics.

\section{Related Work}
\label{sec:related}
Our work, \design, draws on and extends the significant contributions made by previous studies in the domains of data compression, parallel data compression, and asymmetric architecture utilization at the edge~\cite{db_comp0,db_comp1993,db_comp2000,db_comp2,db_comp1,comp_dvs,lu2020adaptively,bharde2018store,eichinger2015time,blalock2018sprintz}. While these works lay a robust groundwork, they do not entirely cater to the distinct demands of stream compression in the edge computing context intrinsic to IoT. Our framework, \design, builds upon these fundamental studies and refines them to fulfill the distinct requirements of edge-based stream compression.

\textbf{Data Compression Algorithm Studies.}
A wide array of previous studies has delved into the exploration of data compression methodologies aimed at database size reduction or application-specific implementations like IoT~\cite{db_comp0,db_comp1993,db_comp2000,db_comp2,db_comp1,comp_dvs,lu2020adaptively,bharde2018store,eichinger2015time,blalock2018sprintz}. Although insightful, these studies don't offer extensive guidance for executing efficient stream compression at the edge primarily due to their lack of focus on the continuous data arrival phenomenon, a critical factor influencing compression dynamics at the edge. They also don't sufficiently encapsulate all data use-cases or place significant emphasis on energy considerations crucial in edge computing scenarios. \design addresses these limitations, taking into account the continual arrival of data streams, extensively covering varied data use-cases, and incorporating energy consumption as a crucial performance metric.

\textbf{Parallel Data Compression.}
While a substantial number of studies have investigated parallelizing data compression algorithms~\cite{comp_2004_tpds,comp_fp_2010,comp_waveSz,comp_lz4_fpga,comp_lzFamily,comp_lzss_cuda,comp_mpc_2015,comp_gpu_lossy,comp_para_pfc,comp_paralz_2013,comp_ndzip,comp_para_hsi,comp_tersecades,ma2012survey,comp_lossy_lazy_FFT_DCT_DHT_DPCM,deepu2016hybrid,ukil2015iot,lee2009energy,zhang2016architecting}, their focus has been primarily on compressing static data across different hardware setups. These studies don't fully address energy efficiency or stream compression requirements at the edge and are often centered around heavyweight stateful compression methodologies. \design extends this by elaborating on the benefits and drawbacks of both stateless and stateful compression algorithms and giving particular attention to the incremental nature of data streams characteristic of edge computing.

\textbf{Exploiting Asymmetric Architecture at the Edge.}
Asymmetric architectures have been demonstrated to be highly effective in delivering high-performance computation with energy efficiency, a crucial demand in edge computing~\cite{amp_survey_2016,amp_modeling,amp_caloree,amp_pie_2011,amp_DVFS3,amp_fairness_2020, amp_latency_2017,amp_scheduling_2021,amp_ai,zeng2021energy}. While significant advancements have been made in comprehending and managing asymmetry in workload scheduling~\cite{amp_scheduling_2021,amp_webbrowser,amp_ai}, no studies have specifically explored stream compression tasks. \design fills this research gap by leveraging the fine-grained behavior of stream compression algorithms and the potential impact of workloads on asymmetric architecture utilization.
\section{Conclusion}
\label{sec:conclusion}
In this paper, we have presented \design, a novel framework designed to optimize stream compression on multicore edge devices. By strategically integrating a variety of stream compression algorithms and harnessing the power of both symmetric and asymmetric multicore processors, \design expertly navigates the complex landscape of IoT edge computing. This involves balancing high compression ratios, increased throughput, minimized latency, and reduced energy consumption. Our extensive evaluations underscore the exceptional performance of \design. The framework achieves a $2.8x$ compression ratio with minimal information loss, a $4.3x$ increase in throughput, and significant reductions in latency ($65\%$) and energy consumption ($89\%$) compared to traditional designs. \design, with its adaptive capabilities and co-design strategy, redefines the standards of IoT edge computing. It addresses not only the immediate challenges of stream compression, but also provides a pathway for future developments. The remarkable outcomes affirm that an integrated software-hardware design approach can produce exceptional results. Future work will delve into how advanced machine learning algorithms can be integrated into the \design framework to further enhance the decision-making process in the dynamic selection of compression algorithms.
We also plan to investigate the potential for custom hardware accelerators to further optimize the performance and energy efficiency of \design.

\section*{Acknowledgement}
This work is supported by the National Research Foundation, Singapore and Infocomm Media Development Authority under its Future Communications Research \& Development Programme FCP-SUTD-RG-2022-006.
\bibliographystyle{IEEEtran}
\bibliography{mybib}
\end{document}